\input harvmac

\input amssym
\input epsf


\newfam\frakfam
\font\teneufm=eufm10
\font\seveneufm=eufm7
\font\fiveeufm=eufm5
\textfont\frakfam=\teneufm
\scriptfont\frakfam=\seveneufm
\scriptscriptfont\frakfam=\fiveeufm
\def\frak{\fam\eufmfam \teneufm}


\def\bb{
\font\tenmsb=msbm10
\font\sevenmsb=msbm7
\font\fivemsb=msbm5
\textfont1=\tenmsb
\scriptfont1=\sevenmsb
\scriptscriptfont1=\fivemsb
}



\newfam\dsromfam
\font\tendsrom=dsrom10
\textfont\dsromfam=\tendsrom
\def\ds{\fam\dsromfam \tendsrom}


\newfam\mbffam
\font\tenmbf=cmmib10
\font\sevenmbf=cmmib7
\font\fivembf=cmmib5
\textfont\mbffam=\tenmbf
\scriptfont\mbffam=\sevenmbf
\scriptscriptfont\mbffam=\fivembf


\newfam\mbfcalfam
\font\tenmbfcal=cmbsy10
\font\sevenmbfcal=cmbsy7
\font\fivembfcal=cmbsy5
\textfont\mbfcalfam=\tenmbfcal
\scriptfont\mbfcalfam=\sevenmbfcal
\scriptscriptfont\mbfcalfam=\fivembfcal


\newfam\mscrfam
\font\tenmscr=rsfs10
\font\sevenmscr=rsfs7
\font\fivemscr=rsfs5
\textfont\mscrfam=\tenmscr
\scriptfont\mscrfam=\sevenmscr
\scriptscriptfont\mscrfam=\fivemscr
\def\scr{\fam\mscrfam \tenmscr}




\def\tilde{\widetilde}
\def\t{\tilde}
\def\hat{\widehat}
\def\h{\hat}
\def\bar{\overline}
\def\b{\bar}
\def\bsq#1{{{\b{#1}}^{\lower 2.5pt\hbox{$\scriptstyle 2$}}}}
\def\bexp#1#2{{{\b{#1}}^{\lower 2.5pt\hbox{$\scriptstyle #2$}}}}
\def\dotexp#1#2{{{#1}^{\lower 2.5pt\hbox{$\scriptstyle #2$}}}}


\def\rt2{\sqrt{2}}
\def\half {{1 \over 2}}
\def\Re{\mathop{\rm Re}}
\def\Im{\mathop{\rm Im}}
\def\d{\partial}

\def\Tr{\mathop{\rm Tr}}


\font\tenbifull=cmmib10
\font\tenbimed=cmmib7
\font\tenbismall=cmmib5
\textfont9=\tenbifull \scriptfont9=\tenbimed
\scriptscriptfont9=\tenbismall

\mathchardef\bbGamma="7000
\mathchardef\bbDelta="7001
\mathchardef\bbPhi="7002
\mathchardef\bbAlpha="7003
\mathchardef\bbXi="7004
\mathchardef\bbPi="7005
\mathchardef\bbSigma="7006
\mathchardef\bbUpsilon="7007
\mathchardef\bbTheta="7008
\mathchardef\bbPsi="7009
\mathchardef\bbOmega="700A
\mathchardef\bbalpha="710B
\mathchardef\bbbeta="710C
\mathchardef\bbgamma="710D
\mathchardef\bbdelta="710E
\mathchardef\bbepsilon="710F
\mathchardef\bbzeta="7110
\mathchardef\bbeta="7111
\mathchardef\bbtheta="7112
\mathchardef\bbiota="7113
\mathchardef\bbkappa="7114
\mathchardef\bblambda="7115
\mathchardef\bbmu="7116
\mathchardef\bbnu="7117
\mathchardef\bbxi="7118
\mathchardef\bbpi="7119
\mathchardef\bbrho="711A
\mathchardef\bbsigma="711B
\mathchardef\bbtau="711C
\mathchardef\bbupsilon="711D
\mathchardef\bbphi="711E
\mathchardef\bbchi="711F
\mathchardef\bbpsi="7120
\mathchardef\bbomega="7121
\mathchardef\bbvarepsilon="7122
\mathchardef\bbvartheta="7123
\mathchardef\bbvarpi="7124
\mathchardef\bbvarrho="7125
\mathchardef\bbvarsigma="7126
\mathchardef\bbvarphi="7127



\def\ibar{\b{i}}
\def\jbar{\b{j}}


\def\jbar{\b{j}}


\def\CF{{\cal F}}

\def\CH{{\cal H}}

\def\CJ{{\cal J}}
\def\CK{{\cal K}}
\def\CL{{\cal L}}
\def\CM{{\cal M}}
\def\CN{{\cal N}}
\def\CO{{\cal O}}

\def\CR{{\cal R}}
\def\CS{{\cal S}}

\def\CV{{\cal V}}


\def\1{{\ds 1}}
\def\R{\hbox{$\bb R$}}
\def\C{\hbox{$\bb C$}}

\def\Z{\hbox{$\bb Z$}}

\def\P{\hbox{$\bb P$}}


\def\fz{{z } 
}
\def\fr{{ r} 
}

\def\bz{{\b z}}

\def\zHt{ \left(\t z-{r\over 2} \t\CH\right) }
\def\zH{ \left( z-{r\over 2} \CH\right) }
\def\jR{j^{(R)}}

\def\gen{{\bf g}}
\def\epsdef{{\epsilon_\Omega}}

\def\tp{\theta^+}
\def\ttp{\t \theta^+}
\def\tm{\theta^-}
\def\ttm{\t \theta^-}
\def\tfour{\tp\tm \ttp\ttm}


\noblackbox

\def\unit{\relax{\rm 1\kern-.26em I}}
\def\nada{\relax{\rm 0\kern-.30em l}}
\def\tilde{\widetilde}


\noblackbox
\def\IL{\relax{\rm I\kern-.18em L}}
\def\IH{\relax{\rm I\kern-.18em H}}
\def\IR{\relax{\rm I\kern-.18em R}}
\def\IC{\relax\hbox{$\inbar\kern-.3em{\rm C}$}}
\def\IZ{\relax\ifmmode\mathchoice
{\hbox{\cmss Z\kern-.4em Z}}{\hbox{\cmss Z\kern-.4em Z}} {\lower.9pt\hbox{\cmsss Z\kern-.4em Z}}
{\lower1.2pt\hbox{\cmsss Z\kern-.4em Z}}\else{\cmss Z\kern-.4em Z}\fi}
\def\CM {{\cal M}}
\def\CN {{\cal N}}
\def\CR {{\cal R}}

\def\CF {{\cal F}}
\def\CJ {{\cal J}}
\def\partialslash{\not{\hbox{\kern-2pt $\partial$}}}

\def\CL {{\cal L}}
\def\CV {{\cal V}}
\def\CO {{\cal O}}

\def\CH {{\cal H}}

\def\CS {{\cal S}}

\def\CK{{\cal K}}
\def\CM {{\cal M}}
\def\CN {{\cal N}}

\def\CO {{\cal O}}

\def\CV{{\cal V }}

\def\CS {{\cal S }}

\def\Tr{{\rm Tr}}

\font\manual=manfnt \def\dbend{\lower3.5pt\hbox{\manual\char127}}

\def\IZ{\relax\ifmmode\mathchoice
{\hbox{\cmss Z\kern-.4em Z}}{\hbox{\cmss Z\kern-.4em Z}} {\lower.9pt\hbox{\cmsss Z\kern-.4em Z}}
{\lower1.2pt\hbox{\cmsss Z\kern-.4em Z}}\else{\cmss Z\kern-.4em Z}\fi}
\def\half {{1\over 2}}

\def\bar{\overline}
\def\CS{{\cal S}}
\def\CH{{\cal H}}

\def\lambdaa{\lambda_{\rm ax}}
\def\lambdav{\lambda_{\rm vec}}

\def\rt2{\sqrt{2}}
\def\irt2{{1\over\sqrt{2}}}

\def\hat{\widehat}
\def\slashchar#1{\setbox0=\hbox{$#1$}           
   \dimen0=\wd0                                 
   \setbox1=\hbox{/} \dimen1=\wd1               
   \ifdim\dimen0>\dimen1                        
      \rlap{\hbox to \dimen0{\hfil/\hfil}}      
      #1                                        
   \else                                        
      \rlap{\hbox to \dimen1{\hfil$#1$\hfil}}   
      /                                         
   \fi}

\def\foursqr#1#2{{\vcenter{\vbox{
    \hrule height.#2pt
    \hbox{\vrule width.#2pt height#1pt \kern#1pt
    \vrule width.#2pt}
    \hrule height.#2pt
    \hrule height.#2pt
    \hbox{\vrule width.#2pt height#1pt \kern#1pt
    \vrule width.#2pt}
    \hrule height.#2pt
        \hrule height.#2pt
    \hbox{\vrule width.#2pt height#1pt \kern#1pt
    \vrule width.#2pt}
    \hrule height.#2pt
        \hrule height.#2pt
    \hbox{\vrule width.#2pt height#1pt \kern#1pt
    \vrule width.#2pt}
    \hrule height.#2pt}}}}
\def\psqr#1#2{{\vcenter{\vbox{\hrule height.#2pt
    \hbox{\vrule width.#2pt height#1pt \kern#1pt
    \vrule width.#2pt}
    \hrule height.#2pt \hrule height.#2pt
    \hbox{\vrule width.#2pt height#1pt \kern#1pt
    \vrule width.#2pt}
    \hrule height.#2pt}}}}
\def\sqr#1#2{{\vcenter{\vbox{\hrule height.#2pt
    \hbox{\vrule width.#2pt height#1pt \kern#1pt
    \vrule width.#2pt}
    \hrule height.#2pt}}}}

\def\figin{\epsfcheck\figin}\def\figins{\epsfcheck\figins}
\def\epsfcheck{\ifx\epsfbox\UnDeFiNeD
\message{(NO epsf.tex, FIGURES WILL BE IGNORED)}
\gdef\figin##1{\vskip2in}\gdef\figins##1{\hskip.5in}
\else\message{(FIGURES WILL BE INCLUDED)}%
\gdef\figin##1{##1}\gdef\figins##1{##1}\fi}
\def\DefWarn#1{}
\def\figinsert{\goodbreak\midinsert}
\def\ifig#1#2#3{\DefWarn#1\xdef#1{fig.~\the\figno}
\writedef{#1\leftbracket fig.\noexpand~\the\figno}%
\figinsert\figin{\centerline{#3}}\medskip\centerline{\vbox{\baselineskip12pt \advance\hsize by
-1truein\noindent\footnotefont{\bf Fig.~\the\figno:\ } \it#2}}
\bigskip\endinsert\global\advance\figno by1}

\lref\FestucciaWS{
  G.~Festuccia and N.~Seiberg,
  ``Rigid Supersymmetric Theories in Curved Superspace,''
JHEP {\bf 1106}, 114 (2011).
[arXiv:1105.0689 [hep-th]].
}

\lref\DumitrescuIU{
  T.~T.~Dumitrescu and N.~Seiberg,
  ``Supercurrents and Brane Currents in Diverse Dimensions,''
  JHEP {\bf 1107}, 095 (2011)
  [arXiv:1106.0031 [hep-th]].
}

\lref\ClossetRU{
  C.~Closset, T.~T.~Dumitrescu, G.~Festuccia and Z.~Komargodski,
  ``Supersymmetric Field Theories on Three-Manifolds,''
[arXiv:1212.3388 [hep-th]].
}

\lref\DumitrescuHA{
  T.~T.~Dumitrescu, G.~Festuccia and N.~Seiberg,
  ``Exploring Curved Superspace,''
JHEP {\bf 1208}, 141 (2012).
[arXiv:1205.1115 [hep-th]].
}

\lref\DumitrescuAT{
  T.~T.~Dumitrescu and G.~Festuccia,
  ``Exploring Curved Superspace (II),''
JHEP {\bf 1301}, 072 (2013).
[arXiv:1209.5408 [hep-th]].
}

\lref\KlareDKA{
  C.~Klare and A.~Zaffaroni,
  ``Extended Supersymmetry on Curved Spaces,''
JHEP {\bf 1310}, 218 (2013).
[arXiv:1308.1102 [hep-th]].
}

\lref\ClossetVRA{
  C.~Closset, T.~T.~Dumitrescu, G.~Festuccia and Z.~Komargodski,
  ``The Geometry of Supersymmetric Partition Functions,''
JHEP {\bf 1401}, 124 (2014).
[arXiv:1309.5876 [hep-th]].
}

\lref\BeniniUI{
  F.~Benini and S.~Cremonesi,
  ``Partition functions of N=(2,2) gauge theories on $S^2$ and vortices,''
[arXiv:1206.2356 [hep-th]].
}

\lref\DoroudXW{
  N.~Doroud, J.~Gomis, B.~Le Floch and S.~Lee,
  ``Exact Results in D=2 Supersymmetric Gauge Theories,''
[arXiv:1206.2606 [hep-th]].
}

\lref\GomisWY{
  J.~Gomis and S.~Lee,
  ``Exact Kahler Potential from Gauge Theory and Mirror Symmetry,''
JHEP {\bf 1304}, 019 (2013).
[arXiv:1210.6022 [hep-th]].
}

\lref\WittenYC{
  E.~Witten,
  ``Phases of N=2 theories in two-dimensions,''
Nucl.\ Phys.\ B {\bf 403}, 159 (1993).
[hep-th/9301042].
}

\lref\ImamuraUW{
  Y.~Imamura,
  ``Relation between the 4d superconformal index and the $S^3$ partition function,''
JHEP {\bf 1109}, 133 (2011).
[arXiv:1104.4482 [hep-th]].
}

\lref\ImamuraSU{
  Y.~Imamura and S.~Yokoyama,
  ``Index for three dimensional superconformal field theories with general R-charge assignments,''
JHEP {\bf 1104}, 007 (2011).
[arXiv:1101.0557 [hep-th]].
}

\lref\ImamuraWG{
  Y.~Imamura and D.~Yokoyama,
  ``N=2 supersymmetric theories on squashed three-sphere,''
Phys.\ Rev.\ D {\bf 85}, 025015 (2012).
[arXiv:1109.4734 [hep-th]].
}

\lref\JafferisUN{
  D.~L.~Jafferis,
  ``The Exact Superconformal $R$-symmetry Extremizes Z,''
JHEP {\bf 1205}, 159 (2012).
[arXiv:1012.3210 [hep-th]].
}

\lref\HamaAV{
  N.~Hama, K.~Hosomichi and S.~Lee,
  ``Notes on SUSY Gauge Theories on Three-Sphere,''
JHEP {\bf 1103}, 127 (2011).
[arXiv:1012.3512 [hep-th]].
}

\lref\BeniniNDA{
  F.~Benini, R.~Eager, K.~Hori and Y.~Tachikawa,
  ``Elliptic genera of two-dimensional N=2 gauge theories with rank-one gauge groups,''
Lett.Math.Phys. (2013).
[arXiv:1305.0533 [hep-th]].
}

\lref\DoroudPKA{
  N.~Doroud and J.~Gomis,
  ``Gauge Theory Dynamics and Kahler Potential for Calabi-Yau Complex Moduli,''
[arXiv:1309.2305 [hep-th]].
}

\lref\ShadchinYZ{
  S.~Shadchin,
  ``On F-term contribution to effective action,''
JHEP {\bf 0708}, 052 (2007).
[hep-th/0611278].
}

\lref\GerasimovEB{
  A.~Gerasimov, D.~Lebedev and S.~Oblezin,
  ``Archimedean L-factors and Topological Field Theories I,''
Commun.\ Num.\ Theor.\ Phys.\  {\bf 5}, 57 (2011).
[arXiv:0906.1065 [math.NT]].
}

\lref\DimofteTZ{
  T.~Dimofte, S.~Gukov and L.~Hollands,
  ``Vortex Counting and Lagrangian 3-manifolds,''
Lett.\ Math.\ Phys.\  {\bf 98}, 225 (2011).
[arXiv:1006.0977 [hep-th]].
}

\lref\KlareGN{
  C.~Klare, A.~Tomasiello and A.~Zaffaroni,
  ``Supersymmetry on Curved Spaces and Holography,''
JHEP {\bf 1208}, 061 (2012).
[arXiv:1205.1062 [hep-th]].
}

\lref\CassaniRI{
  D.~Cassani, C.~Klare, D.~Martelli, A.~Tomasiello and A.~Zaffaroni,
  ``Supersymmetry in Lorentzian Curved Spaces and Holography,''
Commun.\ Math.\ Phys.\  {\bf 327}, 577 (2014).
[arXiv:1207.2181 [hep-th]].
}

\lref\WittenXJ{
  E.~Witten,
  ``Topological Sigma Models,''
Commun.\ Math.\ Phys.\  {\bf 118}, 411 (1988).
}

\lref\EguchiVZ{
  T.~Eguchi and S.~-K.~Yang,
  ``N=2 superconformal models as topological field theories,''
Mod.\ Phys.\ Lett.\ A {\bf 5}, 1693 (1990).
}

\lref\PestunRZ{
  V.~Pestun,
  ``Localization of gauge theory on a four-sphere and supersymmetric Wilson loops,''
Commun.\ Math.\ Phys.\  {\bf 313}, 71 (2012).
[arXiv:0712.2824 [hep-th]].
}

\lref\LosevTP{
  A.~Losev, N.~Nekrasov and S.~L.~Shatashvili,
  ``Issues in topological gauge theory,''
Nucl.\ Phys.\ B {\bf 534}, 549 (1998).
[hep-th/9711108].
}

\lref\NekrasovRJ{
  N.~Nekrasov and A.~Okounkov,
  ``Seiberg-Witten theory and random partitions,''
[hep-th/0306238].
}

\lref\SohniusTP{
  M.~F.~Sohnius and P.~C.~West,
  ``An Alternative Minimal Off-Shell Version of N=1 Supergravity,''
Phys.\ Lett.\ B {\bf 105}, 353 (1981).
}

\lref\SohniusFW{
  M.~Sohnius and P.~C.~West,
  ``The Tensor Calculus and Matter Coupling of the Alternative Minimal Auxiliary Field Formulation of $N=1$ Supergravity,''
Nucl.\ Phys.\ B {\bf 198}, 493 (1982).
}

\lref\GatesNR{
  S.~J.~Gates, M.~T.~Grisaru, M.~Rocek and W.~Siegel,
  ``Superspace Or One Thousand and One Lessons in Supersymmetry,''
[hep-th/0108200].
}


\lref\GatesNK{
  S.~J.~Gates, Jr., C.~M.~Hull and M.~Rocek,
  ``Twisted Multiplets and New Supersymmetric Nonlinear Sigma Models,''
Nucl.\ Phys.\ B {\bf 248}, 157 (1984).
}

\lref\BuscherUW{
  T.~Buscher, U.~Lindstrom and M.~Rocek,
  ``New Supersymmetric $\sigma$ Models With {Wess-Zumino} Terms,''
Phys.\ Lett.\ B {\bf 202}, 94 (1988).
}

\lref\LindstromVC{
  U.~Lindstrom, M.~Rocek, I.~Ryb, R.~von Unge and M.~Zabzine,
  ``New N = (2,2) vector multiplets,''
JHEP {\bf 0708}, 008 (2007).
[arXiv:0705.3201 [hep-th]].
}

\lref\LindstromHX{
  U.~Lindstrom, M.~Rocek, I.~Ryb, R.~von Unge and M.~Zabzine,
  ``Nonabelian Generalized Gauge Multiplets,''
JHEP {\bf 0902}, 020 (2009).
[arXiv:0808.1535 [hep-th]].
}

\lref\GatesVE{
  S.~J.~Gates, Jr. and W.~Merrell,
  ``D=2 N=(2,2) Semi Chiral Vector Multiplet,''
JHEP {\bf 0710}, 035 (2007).
[arXiv:0705.3207 [hep-th]].
}

\lref\LindstromZR{
  U.~Lindstrom, M.~Rocek, R.~von Unge and M.~Zabzine,
  ``Generalized Kahler manifolds and off-shell supersymmetry,''
Commun.\ Math.\ Phys.\  {\bf 269}, 833 (2007).
[hep-th/0512164].
}

\lref\FloreaniniAS{
  R.~Floreanini and R.~Jackiw,
  ``Selfdual Fields as Charge Density Solitons,''
Phys.\ Rev.\ Lett.\  {\bf 59}, 1873 (1987).
}


\lref\JiaFOA{
  B.~Jia and E.~Sharpe,
  ``Curvature Couplings in $\CN=(2,2)$ Nonlinear Sigma Models on $S^2$,''
JHEP {\bf 1309}, 031 (2013).
[arXiv:1306.2398 [hep-th]].
}


\lref\ParkNN{
  D.~S.~Park and J.~Song,
  ``The Seiberg-Witten Kahler Potential as a Two-Sphere Partition Function,''
JHEP {\bf 1301}, 142 (2013).
[arXiv:1211.0019 [hep-th]].
}

\lref\JockersDK{
  H.~Jockers, V.~Kumar, J.~M.~Lapan, D.~R.~Morrison and M.~Romo,
  ``Two-Sphere Partition Functions and Gromov-Witten Invariants,''
Commun.\ Math.\ Phys.\  {\bf 325}, 1139 (2014).
[arXiv:1208.6244 [hep-th]].
}

\lref\BeniniNDA{
  F.~Benini, R.~Eager, K.~Hori and Y.~Tachikawa,
  ``Elliptic genera of two-dimensional N=2 gauge theories with rank-one gauge groups,''
Lett.\ Math.\ Phys.\  {\bf 104}, 465 (2014).
[arXiv:1305.0533 [hep-th]].
}

\lref\HalversonEUA{
  J.~Halverson, V.~Kumar and D.~R.~Morrison,
  ``New Methods for Characterizing Phases of 2D Supersymmetric Gauge Theories,''
JHEP {\bf 1309}, 143 (2013).
[arXiv:1305.3278 [hep-th]].
}

\lref\HoriIKA{
  K.~Hori and M.~Romo,
  ``Exact Results In Two-Dimensional (2,2) Supersymmetric Gauge Theories With Boundary,''
[arXiv:1308.2438 [hep-th]].
}

\lref\BeniniXPA{
  F.~Benini, R.~Eager, K.~Hori and Y.~Tachikawa,
  ``Elliptic genera of 2d N=2 gauge theories,''
[arXiv:1308.4896 [hep-th]].
}

\lref\BFGK{
H.~Baum, T.~Friedrich, R.~Grunewald, I.~Kath,
``Twistor and Killing spinors on Riemannian manifolds,''
 Teubner-Texte zur Mathematik, Band 124, TeubnerVerlag Stuttgart/Leipzig 1991.
}

\lref\CrichignoAA{
  P.~M.~Crichigno,
  ``The Semi-Chiral Quotient, Hyperkahler Manifolds and T-Duality,''
JHEP {\bf 1210}, 046 (2012).
[arXiv:1112.1952 [hep-th]].
}

\lref\Efratwip{
E.~Gerchkovitz, work in progress.
}

\lref\GerchkovitzGTA{
  E.~Gerchkovitz, J.~Gomis and Z.~Komargodski,
  ``Sphere Partition Functions and the Zamolodchikov Metric,''
[arXiv:1405.7271 [hep-th]].
   }

\lref\AdamsVW{
  A.~Adams, H.~Jockers, V.~Kumar and J.~M.~Lapan,
  ``N=1 Sigma Models in $AdS_4$,''
JHEP {\bf 1112}, 042 (2011).
[arXiv:1104.3155 [hep-th]].
}

\lref\HellermanMV{
  S.~Hellerman, D.~Orlando and S.~Reffert,
  ``String theory of the Omega deformation,''
JHEP {\bf 1201}, 148 (2012).
[arXiv:1106.0279 [hep-th]].
}

\lref\KuzenkoUYA{
  S.~M.~Kuzenko, U.~Lindstrom, M.~Rocek, I.~Sachs and G.~Tartaglino-Mazzucchelli,
  ``Three-dimensional N=2 supergravity theories: From superspace to components,''
Phys.\ Rev.\ D {\bf 89}, 085028 (2014).
[arXiv:1312.4267 [hep-th]].
}

\lref\BonelliMMA{
  G.~Bonelli, A.~Sciarappa, A.~Tanzini and P.~Vasko,
  ``Vortex partition functions, wall crossing and equivariant Gromov-Witten invariants,''
[arXiv:1307.5997 [hep-th]].
}


\rightline{ IMPERIAL-TP-14-SC-03}
\vskip-20pt
\Title{
} {\vbox{\centerline{Comments on $\CN=(2,2)$ Supersymmetry}
\vskip2pt
\centerline{ on Two-Manifolds}}}

\vskip-20pt
\centerline{Cyril Closset$^1$ and Stefano Cremonesi$^{2}$}
\vskip15pt
\centerline{ $^{1}$ {\it  Simons Center for Geometry and Physics, Stony Brook University }}
\vskip-5pt
  \centerline{\it    Stony Brook, NY 11794, USA}
 \centerline{$^{2}$ {\it Theoretical Physics Group, Imperial College London}}
\vskip-5pt
  \centerline{\it Prince Consort Road, London SW7 2AZ, UK}

\vskip45pt

\noindent We study curved-space rigid supersymmetry for two-dimensional $\CN=(2,2)$ supersymmetric fields theories with a vector-like $R$-symmetry by coupling such theories to background supergravity. The associated Killing spinors can be viewed as holomorphic sections of particular complex line bundles over Euclidean space-time, which severely restricts the allowed supersymmetric couplings on compact orientable Riemann surfaces without boundaries. For genus ${\bf g}>1$, the only consistent non-singular couplings are the ones dictated by the topological $A$-twist. On spaces with $S^2$ topology, there exist additional supersymmetric backgrounds with $m=0$ or $\pm 1$ unit of flux for the $R$-symmetry gauge field. The $m=-1$ case includes the $\Omega$-background on the sphere. We also systematically work out the curved-space supersymmetry multiplets and supersymmetric Lagrangians.

\vskip10pt

\Date{April 2014}

\listtoc\writetoc
\newsec{Introduction}
There has been some progress recently in the study of two-dimensional $\CN=(2,2)$ supersymmetric field theories, and of supersymmetric gauge theories in particular, using localization techniques ---see for instance \refs{\DoroudXW\BeniniUI\GomisWY\JockersDK\ParkNN\BonelliMMA\BeniniNDA\BeniniXPA\HoriIKA-\DoroudPKA}. A crucial ingredient in this line of development is a proper understanding of rigid supersymmetry on curved space. 
In this paper we revisit the problem of defining $\CN=(2,2)$ supersymmetric theories on any  two-manifold $\Sigma$ with an arbitrary non-dynamical Riemannian metric. 
We will mostly focus on $\Sigma$ a compact orientable Riemann surface without boundaries. We implement the approach to curved-space rigid supersymmetry advocated by Festuccia and Seiberg \FestucciaWS\ (see also \AdamsVW\ for an earlier discussion in the case of $AdS_4$), which realizes any supersymmetric geometry as a supersymmetric background for an off-shell supergravity multiplet. See \refs{\DumitrescuHA\KlareGN\CassaniRI\DumitrescuAT\ClossetRU\KlareDKA\ClossetVRA-\KuzenkoUYA} for related works in higher dimensions.

We consider $\CN=(2,2)$ theories with a vector-like $R$-symmetry. On $\R^2$, the $\CN=(2,2)$ supersymmetry algebra reads
\eqn\susyflat{
\{Q_-, \t Q_-\}= 4 P_z~, \quad \{Q_+, \t Q_+\}= -4 P_\bz~, \quad
\{Q_-, \t Q_+\}= Z~, \quad \{Q_+, \t Q_-\}= \t Z~,
}
with the other anticommutators vanishing. Here $P_z, P_\bz$ are the left- and right-moving momenta respectively, and $Z, \t Z$ is the complex central charge that commutes with the vector $R$-symmetry ($Z$ and $\t Z$ are complex conjugate of each other in Lorentzian signature). The vector-like $R$-charge of the supercharges $Q_\pm$ and $\t Q_\pm$ is $-1$ and $+1$, respectively.
The supersymmetry current sits together with the energy-momentum tensor in a supersymmetry multiplet called the $\CR$-multiplet, which was studied in detail in \DumitrescuIU. The bottom component of the $\CR$-multiplet is the conserved $R$-symmetry current $j_\mu^{(R)}$. The $\CR$-multiplet naturally couples to an off-shell supergravity multiplet which includes an 
$R$-symmetry gauge field $A_\mu$ \GatesNR.

Note that we do not require an axial-like $R$-symmetry. Even when present in flat space, the axial $R$-symmetry is typically broken by the curved space background for massive theories. When the field theory of interest flows to an interacting fixed point, the axial symmetry is restored as an accidental symmetry in the infrared (IR) and the background fields that break it explicitly only couple to redundant operators%
\foot{Redundant operators of a CFT are operators that vanish in any correlation function at separated points.} of the IR superconformal field theory. (In this paper, whenever we write $R$-symmetry we will mean the vector $R$-symmetry unless otherwise stated.)

It is well-known that one can preserve supersymmetry on any orientable Riemann surface by the so-called topological $A$-twist \WittenXJ, which corresponds to identifying $A_\mu$ with the spin connection or redefining the energy-momentum tensor \EguchiVZ.%
\foot{Similarly, the topological $B$-twist, which is done with respect to an axial $R$-symmetry, can be obtained by coupling the two-dimensional Ferrara-Zumino (FZ) multiplet (see for instance \DumitrescuIU) to its corresponding background supergravity. By the $\Z_2$ mirror action, this is equivalent to considering the $A$-twist and replacing superfields by their twisted counterparts ({\it e.g.} chiral multiplets by twisted chiral multiplets, and vector multiplets by twisted vector multiplets).} One natural question that this work answers is whether there are more general ways to preserve supersymmetry on $\Sigma$. We will see that on Riemann surfaces of genus greater than one there are no other possibilities than the $A$-twist ---perhaps unsurprisingly given that higher genus Riemann surfaces do not admit ordinary Killing spinors \BFGK. The case of genus one will be omitted from our discussion because the supersymmetry is essentially the same as in flat space. The most interesting supersymmetric backgrounds occur at genus zero. There are three topologically distinct ways to preserve some supercharge(s) on the sphere, corresponding to having $-1$, $0$ or $1$ unit of flux for the $R$-symmetry gauge field. The case of vanishing flux has been studied recently in \refs{\DoroudXW,\BeniniUI,\GomisWY}, which computed the partition functions of rather generic $\CN=(2,2)$ gauge theories on such spheres using localization methods. The case of $\mp 1$ unit of flux corresponds to the $A$- (and $\b A$-)twist and to some deformations thereof. In particular, it includes the so-called $\Omega$-background \refs{\NekrasovRJ,\ShadchinYZ} on any $S^2$ with a $U(1)$ isometry.

While our formalism is very similar to its higher dimensional counterparts in three and four dimensions for theories with four supercharges and an $R$-symmetry \refs{\DumitrescuHA,\KlareGN,\ClossetRU}, it is worth pointing out some interesting peculiarities of the two-dimensional setup. First, while supersymmetry implies the presence of a complex structure in four dimensions (or of a transversely holomorphic foliation in three dimensions), any orientable two-manifold $\Sigma$ is complex and the complex structure is not related to the presence of a Killing spinor. The complex structure of  $\Sigma$ will however play an important role because the most convenient way to describe the Killing spinors is in terms of holomorphic sections of some complex line bundles over $\Sigma$.  Another two-dimensional peculiarity is that the elementary Killing spinors (which are Weyl spinors $\zeta_\pm$) have zeros, which does not happen in \refs{\DumitrescuHA,\KlareGN,\ClossetRU}.%
\foot{When we uplift from two to three (or four) dimensions, the 2d Weyl spinors $\zeta_-, \zeta_+$ combine into a three-dimensional Dirac spinor (or four-dimensional Weyl spinor) $\zeta^T=(\zeta_-, \zeta_+)$, which is nowhere vanishing \refs{\DumitrescuHA,\KlareGN,\ClossetRU}.} In this respect, $\CN=(2,2)$ curved-space rigid supersymmetry is closer to $\CN=2$ curved-space supersymmetry in four dimensions (see for instance \refs{\PestunRZ,\KlareDKA}).

This paper is organized as follows. In section 2, we derive the Killing spinor equations governing rigid supersymmetry in two-dimensions by taking the rigid limit of the ``new minimal'' supergravity multiplet. In section 3, we classify regular supergravity backgrounds preserving at least one supercharge. In sections 4 and 5, we discuss the case of two and four supercharges, respectively. In section 6, we present the curved-space rigid supersymmetry algebra and its realization on various supersymmetry multiplets, and we give the curved-space generalization of many standard flat-space supersymmetric Lagrangians. Our conventions are spelled out in Appendix, which also contains various useful formulas. In particular,  we relate our results  to the higher-dimensional results of \refs{\DumitrescuHA,\KlareGN,\ClossetRU}  by dimensional reduction in Appendix C.

\newsec{The Two-Dimensional Killing Spinor Equations}
In this section, we derive the generalized Killing spinor equations which governs curved space supersymmetry for any $\CN=(2,2)$ supersymmetric theory with a vector $R$-symmetry. Following \FestucciaWS,   we couple the $\CR$-multiplet to supergravity and read off the Killing spinor equations from the gravitino variations. The relevant two-dimensional supergravity is the dimensional reduction of the new minimal supergravity in four dimensions \refs{\SohniusTP,\SohniusFW} discussed in \GatesNR. We will only need to consider linearized supergravity, similarly to the analysis of \ClossetRU. 

\subsec{$\CR$-Multiplet and Linearized Supergravity}
In any $\CN=(2,2)$ field theory with a vector-like $R$-symmetry $U(1)_R$, there exists a supercurrent multiplet, called the  $\CR$-multiplet, which contains the $U(1)_R$ conserved current as its lowest component. The $\CR$-multiplet $\CR_\mu$ satisfies \DumitrescuIU  
\eqn\Rmultdefi{\eqalign{
&\t D_+ \CR_z = -{1\over 4} \chi_-~, \quad  \t D_- \CR_{\b z} = -{1\over 4} \chi_+~,
\quad D_+ \CR_z = -{1\over 4} \t\chi_-~,\quad  D_- \CR_{\b z} = -{1\over 4} \t \chi_+~,\cr
& \t D_{\pm }\chi_+=0~, \;\quad\qquad \t D_{\pm }\chi_-= 0~, \qquad\quad\;  D_{\pm }\t\chi_+=0~,  \quad\qquad\;  D_{\pm }\t\chi_-= 0~,\cr
& D_+\chi_- -\t D_- \t \chi_+ =k~, \qquad\qquad\qquad\qquad \t D_+\t\chi_- - D_-  \chi_+ =k~,
}}
where $D_\pm$, $\t D_\pm$ are supercovariant derivatives. (See Appendix A for our flat-space conventions.)
The constant $k$ is a space-filling brane charge \DumitrescuIU. We set $k=0$ in the following. Expanding in components, we have
\eqn\Rmultcompo{\eqalign{
\CR_{z} = \, &  \jR_z - i \theta^- S_{- z}- i \theta^+ S_{+ z}- i\t \theta^- \t S_{- z}- i \t \theta^+\t S_{+ z} 
                         + 4 \tm \ttm T_{zz} - 4 \tp\ttp T_{\b z z}\cr
& - {i\over 2} \tp \ttm  j^{(\t Z)}_z + {i\over 2} \tm\ttp j^{(Z)}_z - 2 \tp\tm\ttm \d_z S_{+z}- 2\tp \tm \ttp \d_\bz S_{-z}\cr
&  - 2 \ttp\ttm\tm \d_z \t S_{+z} - 2\ttp \ttm \tp \d_\bz \t S_{-z} + 4 \tfour \d_z \d_\bz \jR_z \, , \cr
 \CR_\bz = \, &  \jR_z - i \theta^- S_{- \bz}- i \theta^+ S_{+ \bz}- i\t \theta^- \t S_{- \bz}- i \t \theta^+\t S_{+ \bz} 
                         - 4 \tp \ttp T_{\bz\bz} + 4 \tm\ttm T_{z \bz}\cr
& - {i\over 2} \tp \ttm j^{(\t Z)}_\bz + {i\over 2} \tm\ttp j^{(Z)}_\bz - 2 \tp\tm\ttp \d_\bz S_{-\bz}- 2\tp \tm \ttm \d_z S_{+\bz}\cr
&  - 2 \ttp\ttm\tp \d_\bz \t S_{-\bz} - 2\ttp \ttm \tm \d_z \t S_{+\bz} + 4 \tfour \d_z \d_\bz \jR_\bz \, .
}}
Here $\jR_\mu$ is the $R$-symmetry current, $S_{\pm \mu}$,  $\t S_{\pm \mu}$ are the supersymmetry currents and  $j^{(Z)}_\mu,  j^{(\t Z)}_\mu$ are conserved currents for the complex central charge $Z, \t Z$ in \susyflat. The energy momentum tensor $T_{\mu\nu}$ is symmetric and conserved. 
At first order around flat space, the  $\CR$-multiplet couples minimally to a linearized supergravity multiplet $\CH_\mu$:
\eqn\sugracoupling{
{\scr L}_{\rm sugra}= - 4 \int d^4\theta (\CH_z \CR_\bz+\CH_\bz \CR_z)~.
}
Due to the constraints \Rmultdefi, the superfield $(\CH_z, \CH_\bz)$ enjoys the gauge freedom 
\eqn\gaugeinvH{
\delta_L \CH_z = - D_- \t L_- + \t D_- L_-\, , \qquad 
\delta_L \CH_\bz =  D_+ \t L_+ - \t D_+ L_+\, ,
}
where $L_\pm, \t L_\pm$ are fermionic multiplets such that
\eqn\constrR{
\t D_+ D_- D_+ \t L_- - \t D_- D_- D_+ \t L_+ +  D_+ \t D_-\t D_+  L_- -  D_-\t D_-\t D_+  L_+ =0~.
}
One can use \gaugeinvH\ to fix a Wess-Zumino gauge for $\CH_\mu$:
\eqn\CHmu{\eqalign{
\CH_z = &\,   -\tm\ttm h_{z z} + \tp \ttp h_{\bz z} +\half \tm\ttp C_z  + \half \tp \ttm \t C_z 
 + i\tp \tm \ttp \t \psi_{+  z} \cr & + i\tp \tm \ttm \t \psi_{-  z}   - i\ttp \ttm \tp \psi_{+  z}  - i\ttp \ttm \tm \psi_{-  z}  + 2 \tfour  A_z~,\cr
\CH_\bz = &  -\tm\ttm h_{z\bz} + \tp \ttp h_{\bz\bz} +\half\tm\ttp C_\bz  + \half \tp \ttm \t C_\bz 
 + i\tp \tm \ttp \t \psi_{+  \bz} \cr & + i\tp \tm \ttm\t \psi_{-  \bz}   - i\ttp \ttm \tp\psi_{+  \bz}  - i\ttp \ttm \tm \psi_{-  \bz}  + 2 \tfour  A_\bz~,
}} 
with $h_{z\bz}= h_{\bz z}$. The residual gauge transformations are
\eqn\resgauge{\eqalign{
& \delta_L h_{\mu\nu} = \d_\mu \xi_\nu + \d_\nu \xi_\mu~, \qquad  \delta_L C_\mu = \d_\mu \Lambda^{(C)}~, \qquad  \delta_L \t C_\mu = \d_\mu \Lambda^{(\t C)}~,\cr
& \delta_L A_\mu = \d_\mu \Lambda^{(A)}~,\;\;\qquad \qquad   \delta_L \psi_{\pm\mu} = \d_\mu \varepsilon_\pm~, \qquad \delta_L \t\psi_{\pm\mu} = \d_\mu \t\varepsilon_\pm~.
}}
We can therefore identify $h_{\mu\nu}$ with the graviton. The complex gauge fields $C_\mu, \t C_\mu$ are graviphotons which couple to the complex central charges $Z, \t Z$ in the flat-space supersymmetry algebra \susyflat, $A_\mu$ is a gauge field coupling to the vector $R$-symmetry current and $\psi_{\pm\mu}, \t\psi_{\pm\mu}$ are gravitini. In Euclidean signature, the superfields $L$ and $\t L$ are not complex conjugate of each other, and  the gauge parameters in \resgauge\ are generally complex. We will impose that the metric is real, therefore $\xi_\mu$ must be real in \resgauge. The other background fields $C_\mu$, $\t C_\mu$, $A_\mu$ are allowed to take general values, but we will restrict ourselves to real gauge transformations (this means that $\Lambda^{(\t C)}$ is the complex conjugate of $\Lambda^{(C)}$ while $\Lambda^{(A)}$ is real) because the theories we consider are generally only invariant under those real gauge transformations. 
In Wess-Zumino gauge, the linearized supergravity coupling \sugracoupling\ reads
\eqn\LagSug{
{\scr L}_{\rm sugra} 
= - h^{\mu\nu}T_{\mu\nu} + A^\mu j_\mu^{(R)}-{i\over 8}\big(  C^\mu j_\mu^{(\t Z)} -\t C^\mu j_\mu^{(Z)}\big) -\half\big( S^\mu \psi_\mu - \t S^\mu\t\psi\big)~.}
Note that $C_\mu, \t C_\mu$ couples to the conserved current $j_\mu^{(\t Z)}, j_\mu^{(Z)}$, respectively. 
In the following, we will mostly encounter the graviphoton dual field strengths, defined as
\eqn\defHHtfromB{
\CH = -i \epsilon^{\mu\nu}\d_\mu C_\nu~, \qquad  \quad
\t\CH = -i \epsilon^{\mu\nu}\d_\mu \t C_\nu~.
}
Note that the definition \defHHtfromB\ is also valid in curved space (with $\epsilon^{\mu\nu}$ the Levi-Civita tensor). On a compact two-manifold $\Sigma$, the presence of non-zero flux for $A_\mu, C_\mu, \t C_\mu$ will generally lead to quantization conditions for the charges $R, Z, \t Z$. 

\subsec{Gravitino Variations and Killing Spinor Equations}
Curved space supersymmetry for $\CN=(2,2)$  theories with an $\CR$-multiplet is dictated by the rigid limit of the supergravity multiplet introduced above. A supersymmetric background is such that the supersymmetry variations of the gravitini vanish for some non-trivial Killing spinors.  In the linearized theory, we find
\eqn\dQpsiii{\eqalign{
&\delta_Q \psi_{- z} =  2 \epsilon_-(\d_\bz h_{zz}  - \d_z h_{z\bz} - i A_z)~, \cr
&\delta_Q \psi_{+ z} = 2 \epsilon_+(\d_z h_{z\bz} - \d_\bz h_{zz} - i A_z) - \epsilon_-\t\CH~, \cr
&\delta_Q \psi_{- \bz} =   2 \epsilon_-(\d_\bz h_{z\bz}- \d_z h_{\bz \bz} - i A_\bz )-  \epsilon_+ \CH~,\cr
&\delta_Q \psi_{+ \bz} =  2 \epsilon_+(\d_z h_{\bz\bz} - \d_\bz h_{z\bz} - i A_\bz)~,
}}
with $\epsilon_\pm$ a constant spinor. The supersymmetry variations $\delta_{\t Q} \t\psi_{\pm\mu}$ are similar (in terms of a constant spinor $\t\epsilon_\pm$). We can infer the gravitino variations in the full non-linear theory  (in the rigid limit) from \dQpsiii\  by promoting $\epsilon_\pm$ to  a space-dependent supersymmetry parameter  $\zeta_\pm(x)$.
The final answer is completely fixed by diffeomorphism invariance and dimensional analysis. See \ClossetRU\ for more details about this procedure. 

Any orientable two-dimensional manifold $\Sigma$ is complex. Let us introduce a Hermitian metric $g_{\mu\nu}$ on $\Sigma$ and a complex zweibein $e^1= e^1_z dz, e^{\b 1}= e^{\b 1}_\bz d\bz$ such that
\eqn\genmetrici{
ds^2= 2 g_{z\bz}(z, \bz)\, dz d\bz=  e^1 e^{\b 1}~. 
}
We will often use the holomorphic and antiholomorphic frame indices $1, \b 1$ in the following. (See Appendix A for more details.)
The equations $\delta_Q \psi_{\pm \mu}=0$ are equivalent to the following (generalized) Killing spinor equations 
\eqn\KSEzeta{\eqalign{
& (\nabla_z -i A_z) \zeta_- =0~, \qquad \qquad\;\; \qquad\qquad  (\nabla_\bz -i A_\bz) \zeta_- = \half\CH\, e^{\b 1}_\bz\, \zeta_+~,\cr
& (\nabla_z -i A_z)  \zeta_+ = \half\t\CH\, e^{ 1}_z \,\zeta_-~,  \qquad\quad\qquad\;   (\nabla_\bz -i A_\bz) \zeta_+ = 0~.
}}
Similarly, from $ \delta_{\t Q} \t\psi_{\pm \mu}=0$ we obtain
\eqn\KSEzetat{\eqalign{
& (\nabla_z +i A_z) \t\zeta_- =0~, \qquad \qquad\;\; \qquad\qquad  (\nabla_\bz +i A_\bz) \t\zeta_- = \half\t\CH\, e^{\b 1}_\bz\,\t\zeta_+~,\cr
& (\nabla_z +i A_z) \t \zeta_+ = \half\CH\, e^{ 1}_z \,\t\zeta_-~,  \qquad\quad\qquad\;   (\nabla_\bz +i A_\bz) \t\zeta_+ = 0~.
}}
The Killing spinors $\zeta_\pm$ and $\t\zeta_\pm$ carry $R$-charge $1$ and $-1$, respectively, and the Killing spinor equations are invariant under complexified local vector $R$-symmetry transformations. They are also invariant under the axial $R$-symmetry transformations
\eqn\axialR{\eqalign{
&\zeta_-\rightarrow \lambdaa \zeta_-~, \qquad \zeta_+\rightarrow \lambdaa^{-1}\zeta_+~, \qquad
\t\zeta_-\rightarrow \lambdaa^{-1}\zeta_-~, \qquad \t\zeta_+\rightarrow \lambdaa\t\zeta_+~, \cr
&A_\mu \rightarrow A_\mu~, \qquad\quad\CH\rightarrow \lambdaa^2 \CH~,\qquad\;\;
\t\CH\rightarrow \lambdaa^{-2} \t \CH~,
}}
with $\lambdaa$ a complex constant.
Note that \KSEzetat\ can be formally obtained from \KSEzeta\ by
\eqn\replKSEitoii{
\zeta_\pm \rightarrow \t \zeta_\pm~, \qquad A_\mu\rightarrow - A_\mu~, \qquad \CH\rightarrow \t\CH~, \qquad \t\CH\rightarrow \CH~.
}

The equations \KSEzeta, \KSEzetat\ subsume the various Killing spinor equations used in previous works such as \refs{\DoroudXW,\BeniniUI,\GomisWY}, which studied $\CN=(2,2)$ theories for particular supersymmetric background on the two-sphere. 
To make contact with the notation of those papers, let us consider the Dirac spinors $\zeta^T= (\zeta_-, \zeta_+)$, $\t\zeta^T= (\t\zeta_-, \t\zeta_+)$ and the two-dimensional gamma matrices $\gamma^\mu, \gamma^3$ ---see Appendix A  for  our conventions. We also introduce the supergravity background fields $H, G$ such that
\eqn\defHG{
\CH= H + i G~, \qquad\quad \t\CH= H - i G~.
}
For unitary theories, the fields $H, G$ would be real in Lorentzian signature. To preserve supersymmetry in Euclidean signature we must generally give them complex expectation values, which violates reflection positivity \FestucciaWS.
The Killing spinor equations \KSEzeta, \KSEzetat\ read
\eqn\KSERi{\eqalign{
& (\nabla_\mu -i A_\mu) \zeta = -\half H \gamma_\mu \zeta +{i\over 2} G\gamma_\mu \gamma^3 \zeta~,\cr
&(\nabla_\mu +i A_\mu) \t\zeta = -\half H \gamma_\mu\t \zeta -{i\over 2} G\gamma_\mu \gamma^3\t \zeta~.
}}
For instance, \BeniniUI\ considered the round metric on $S^2$ with $H=-{i\over R_{S^2}}$ (with $R_{S^2}$ the $S^2$ radius) and $A_\mu=G=0$. The authors of \DoroudXW\ considered instead a round $S^2$ with $A_\mu=H=0$ and $G= -{i\over R_{S^2}}$. Those two backgrounds are part of a continuous family of backgrounds preserving four supercharges on the round $S^2$, which are all related by \axialR.

\newsec{Backgrounds Preserving One Supercharge}
Consider an  orientable compact Riemann surface $\Sigma_\gen$ of genus $\gen$ with a  Hermitian metric \genmetrici. Let $\sqrt{g}= 2 g_{z\bz}(z, \bz)$ be the square root of the metric determinant in complex coordinates,  which transforms according to 
\eqn\transfosg{
 \sqrt{g}' = U \b U \sqrt{g}~,\qquad {\rm with }\quad U \equiv {\d z\over  \d z'}~,\quad  \b U \equiv {\d \bz\over  \d \bz'}~,
}
under change of holomorphic coordinates. In other words, $\sqrt{g}$ is a global section of the determinant line bundle.  In this section, we classify backgrounds that preserve at least one supercharge, corresponding to solutions to the Killing spinor equations \KSEzeta. 

\subsec{Solving the Killing Spinor Equation}
Locally, given any metric \genmetrici,  one can use \KSEzeta\  to solve for the supergravity background fields in terms of the Killing spinors. If $\zeta_+ \zeta_-\neq 0$, we have
\eqn\backgrfieldsgen{\eqalign{
&A_z= - i \,\d_z \log \left({ \zeta_- \over g^{{1\over 8}}}\right)~, \qquad \qquad \CH = {2\over \sqrt g} \d_\bz \left(g^{1\over 4} {\zeta_-\over\zeta_+}\right)~,\cr
&A_\bz =  - i \,\d_\bz \log\left({ \zeta_+ \over g^{1\over 8}}\right)~, \qquad \qquad  \t\CH = {2\over \sqrt g} \d_z \left(g^{1\over 4}{\zeta_+\over\zeta_-}\right)~.
}}
The case $\zeta_+ \zeta_- =0$ will be discussed separately below. 
Given a globally defined solution $\zeta_\pm$, we can construct the one-form $p_\mu =\zeta \gamma_\mu \zeta$ of $R$-charge $2$. 
Its holomorphic and antiholomorphic components
\eqn\pzpzb{
p_z =  g^{1\over 4}\, (\zeta_-)^2~,\qquad \qquad  p_\bz = - g^{1\over 4} \,(\zeta_+)^2~
}
are sections of $\CK \otimes L^2$ and $\b \CK \otimes L^2$ respectively, where $\CK$ and $\b\CK$ are the canonical and anticanonical line bundles while $L$ denotes the $U(1)_R$ line bundle. The objects \pzpzb\ transform as
\eqn\transfopz{
p_{z'}'  =   (W_R)^2\, U \, p_z~, \qquad p_{\bz'}' = (W_R)^2\, \b U \, p_\bz~, 
}
between coordinate patches, with $U$ defined  in \transfosg\ and $W_R$ a corresponding transition function for  $L$. It follows that $\zeta_\pm$ transform as
\eqn\transfozeta{
\zeta_-' = W_R \left(U \over \b U\right)^{1\over 4} \, \zeta_-~, \qquad\quad
\zeta_+' = W_R \left(U \over \b U\right)^{-{1\over 4}} \, \zeta_+~,
}
under a holomorphic coordinate change.
A supersymmetric background is partly determined by a choice of $U(1)_R$ line bundle $L$, which is equivalent to a choice of transition functions $W_R$. As mentioned above, we should restrict ourselves to real $R$-symmetry gauge transformations, which means that $W_R$ is a pure phase. Topologically, $U(1)$ line bundles are determined by their first Chern class (or degree). Let us denote
\eqn\coneL{
c_1(L) = m \in \Z~.
}
In general, for each $m$ we have a $2\gen$ real-dimensional family of inequivalent line bundles $\Sigma_\gen$, corresponding to turning on Wilson lines. However, Wilson lines for the vector $R$-symmetry would break supersymmetry and are therefore disallowed.%
\foot{When the theory possesses both the vector and axial $R$-symmetries, one could turn on Wilson lines for the left-moving (or right moving) $R$-symmetry, which would still preserve the right-moving (resp. left-moving) supercharges. This is what is done in the case of the $\CN=(2,2)$ elliptic genus (see for instance \BeniniNDA). Such Wilson lines are not available to us because  we only consider the $\CR$-multiplet and its corresponding background supergravity fields.} Without further loss of generality, we take $L$ such that 
\eqn\LfromK{
L_{\C} \cong \sqrt{\CK}^{\otimes n}~, \qquad {\rm with}\qquad n \equiv {m\over \gen -1}~.
} 
Here $L_{\C}$ is  the holomorphic line bundle associated to $L$,%
\foot{Namely, $L_{\C}$ is such that the phase of its transition functions  are the transition functions of the $U(1)$ bundle $L$: If $\alpha(z)$ is a transition function of $L_{\C}$,  the corresponding transition function of $L$ is  $\sqrt{\alpha/ \b\alpha}$.} 
while $\sqrt{\CK}$ corresponds to the  spin bundle with fully periodic fermions. (In general there are many inequivalent ways of taking the $(\gen-1)$-th root of $\sqrt{\CK}$ in \LfromK, but this subtlety will not affect our conclusions.) It follows from \LfromK\ that
\eqn\transitionfunctionsWR{
W_R= \left(U \over \b U\right)^{n\over 4}\, .
}
Note that, if either $\zeta_-$ or $\zeta_+$ is nowhere vanishing, its associated line bundle is trivial and therefore  \transfozeta\ implies that $n=-1$ or $n=1$, respectively.  
To proceed further, we note that the sections $\zeta_\pm$ can be factorized into a globally defined piece which transforms like \transfosg\ and a holomorphic or an antiholomorphic piece. 
This leads to the following ansatz for  $\zeta_\pm$:
\eqn\partialsol{
\zeta_-(z, \b z) = \lambdav(z,\bz) \lambdaa(z,\bz) g^{1+n \over 8}\, \b s_-(\b z)~, \qquad 
\zeta_+(z, \b z) = {\lambdav(z,\bz) \over \lambdaa(z,\bz)} g^{1-n \over 8} \, s_+(z)~.
}
Here $\lambdav$ and $\lambdaa$ are globally defined nowhere vanishing functions on $\Sigma_\gen$, while $\b s_-$ and $s_+$ transform as
\eqn\transfofoneftwobis{
\b s_-(\bz')' = \, {1\over  \b U^{{1+n\over 2}}}  \, \b s_-(\bz)~, \qquad\quad
s_+(z')' = \, {1\over   U^{{1-n\over 2}}}  \, s_+(z)~.
}
 In other words, $s_\pm$ are identified as {\it (locally) holomorphic sections} of some power of the spin bundle $\sqrt{\CK}$:
\eqn\sectionsf{
s_- \in \Gamma\left(\sqrt{\CK }^{{\,-1-n}} \right)~, \qquad\quad s_+\in \Gamma\left( \sqrt{ \CK }^{{\,-1+n}} \right)~.
}
Plugging \partialsol\ into \backgrfieldsgen, we find
\eqn\backsol{\eqalign{
&A_z={n\over 2} \omega_z - i \,\d_z \log \left(\lambdav \lambdaa \right)~, \qquad \qquad \CH ={2\over \sqrt g} \d_\bz \left( g^{1+ n\over 4}{ \lambdaa^2 \, \b s_- \over s_+}\right)~,\cr
&A_\bz ={n\over 2} \omega_\bz - i \,\d_\bz \log \left({\lambdav \over \lambdaa }\right)~, \qquad \qquad  \t\CH= {2\over \sqrt g}  \d_z \left( g^{1-n \over 4} {s_+\over \lambdaa^2\, \b s_-}\right)~,
}}
with $\omega_\mu$ the spin connection. Note that $\lambdav$ corresponds to a vector $R$-symmetry pure gauge transformation, while $\lambdaa$ can be interpreted  as a local axial $R$-symmetry transformation (recall however that there is no gauge connection for the axial $R$-symmetry). Using the Gauss-Bonnet theorem, one can check that
\eqn\FofAc{
c_1(L) = {1\over 2\pi} \int_\Sigma dA  =  n(\gen-1) = m~,
}
as it should be by construction.

The careful reader will have noted that several different factorizations would have been possible in \partialsol. For instance we could factorize a holomorphic piece in $\zeta_-$ instead of an antiholomorphic one. However,  such alternate factorizations introduce physical poles in $A_\mu$ when plugged into \backgrfieldsgen, corresponding to the zeros (or poles) of the corresponding holomorphic (or meromorphic) sections. We do not consider such singular supergravity backgrounds.

\subsec{The $A$-Twist}
The special cases when either $\zeta_+$ or $\zeta_-$ identically vanishes must be treated separately. For $\zeta_-=0$, the general solution to  \KSEzeta\ reads
\eqn\backgrfieldsAtwist{
A_\mu = \half \omega_\mu - i \, \d_\mu \log \zeta_+~, \qquad \quad \CH =0~,
 \qquad \quad   \t\CH = ({\rm arbitrary})~.
}
This corresponds to a line bundle with $n=1$ in \transitionfunctionsWR. Then $\zeta_+$ is effectively a scalar --- a section of a trivial line bundle --- and we can set it to a constant by a complexified $U(1)_R$ gauge transformation.  Similarly, for $\zeta_+=0$ we have 
\eqn\backgrfieldsAbartwist{
A_\mu = -\half \omega_\mu - i\, \d_\mu \log \zeta_-~, \qquad \quad \CH = ({\rm arbitrary})~,
 \qquad \quad   \t\CH =0~,
}
with $n=-1$. The backgrounds \backgrfieldsAtwist\ and \backgrfieldsAbartwist\  correspond to the $A$- and $\b A$-twists, respectively. (More precisely, this is called the $\half A$-twist when only one supercharge is involved. However this background also preserves a second supercharge of opposite $R$-charge, which we will consider in the next section.) The possibility to turn on an arbitrary background for $\CH$ or $\t\CH$ does not affect the supersymmetry algebra.

\subsec{Global Properties of the Killing Spinors}
Since we are considering compact Riemann surfaces without punctures, the sections $s_\pm$ in \sectionsf\ should be globally holomorphic, so that supersymmetry maps smooth field configurations into smooth field configurations. Since there are no holomorphic sections of a line bundle of negative degree over $\Sigma_\gen$, this regularity condition severely restricts the allowed Killing spinors. Moreover, one should verify that the background fields $\CH, \t \CH$ determined by \backsol\ have no singularities at the zeros of $s_\pm$.

The canonical bundle of a Riemann surface of genus $\gen$ has degree $2\gen-2$, therefore the line bundles with sections \sectionsf\ have degrees $(\gen-1)(-1 \mp n)= 1-\gen\mp m$. For  $\gen>1$, at least one of the two line bundles has negative degree and no holomorphic section. Therefore either $s_-$ or $s_+$ is identically zero, and the only supersymmetric backgrounds correspond to the well-known $A$- or $\b A$-twist \backgrfieldsAtwist\ or \backgrfieldsAbartwist.

On the torus ($\gen=1$), the spinors $\zeta_\pm$ are sections of a trivial line bundle and the supersymmetry is essentially the same as in flat space. We will not study the torus in any detail in this paper. 

The genus zero case is somewhat richer. On the Riemann sphere, \sectionsf\ reads
\eqn\sectionsfCP{
s_- \in \Gamma\left(\CO(n+1) \right)~, \qquad s_+\in \Gamma\left(\CO(-n+1) \right)~,
}
in the standard notation $\CO(k) \cong H^{\otimes k}$ for $\C\P^1$. Regularity imposes that $|n|\leq 1$ for any non-trivial solution. The case $n = \pm 1$ leads to the $A$-twist mentioned above and to some deformation thereof. The case $n=0$ (with no flux for the $R$-symmetry) corresponds to the supersymmetric spheres of \refs{\DoroudXW, \BeniniUI,\GomisWY} and their generalization to more general metrics on the sphere.

\subsec{Flux and Charge Quantization}
The  presence of non-trivial flux for the gauge fields in the supergravity multiplet can lead to restrictions on the allowed values of the charges $R, Z, \t Z$. The  $R$-charge has the standard Dirac quantization condition,
\eqn\Rquantized{
r m \in \Z~, \qquad m\equiv {1\over 2\pi }\int_\Sigma dA~,
}
with $r$ the $R$-charge and $m$ the flux from the real part of $A_\mu$. (The imaginary part of $A_\mu$ is a well-defined one-form by assumption, therefore it does not lead to any flux.)  Similarly, the restriction on $Z, \t Z$ due to non-trivial fluxes for the graviphotons reads
\eqn\quantZ{ \Re {1\over 4\pi} \int_\Sigma d^2 x \sqrt{g} \,\left(z \t \CH - \t z \CH\right) \in \Z~,}
with $z, \t z$ the central charges.

\subsec{The Case of  $\C\P^1$}
Let us consider supersymmetry on the Riemann sphere in more details. We can cover the sphere with two patches with coordinates $z$ and $z'={1\over z}$, respectively, and consider an arbitrary metric \genmetrici. 
In the case of vanishing flux for the $R$-symmetry gauge field, $n=m=0$, we have the Killing spinor
\eqn\partialsol{
\zeta = \left(\matrix{ \zeta_-\cr \zeta_+ }\right) = g^{1 \over 8} \left(\matrix{ \lambdaa\,  \b s_-(\b z) \cr   \lambdaa^{-1} \, s_+(z)}\right)~,
}
up to a complexified $R$-symmetry gauge transformation. (We write all quantities in the northern patch with complex coordinate $z$.) Here $s_\pm$ are both holomorphic sections of $\CO(1)$, which therefore have a single zero on the sphere. Since the zeros of $s_-$ and $s_+$ do not coincide, we can set $s_- =1$ and $s_+= z$ by a change of coordinates. (Then $s_-$  has a zero at the ``south pole'' $z'=0$ while $s_+$ has a zero at the ``north pole'' $z=0$.) A necessary and sufficient condition for the background fields $\CH, \t\CH$ given by \backgrfieldsgen\ to be regular is that the metric takes the form
\eqn\metricnearzero{
g_{z\bz} \sim c_0 + c_1 |z|^2~, \qquad
g_{z'\bz'} \sim c_2 + c_3 |z'|^2~,\qquad c_0, c_1, c_2,c_3\in\R~,
}
near the zeros $z=0$ and $z'=0$, respectively, and similarly for the function $\lambdaa$.%
\foot{It follows from the analysis of the next section that we then have a second supercharge $\t\zeta_\pm$ locally, near the zeros of $s_\pm$.} In the case of $n=-m=1$ unit of flux for the $R$-symmetry, we have the Killing spinor
\eqn\partialsol{
\zeta = \left(\matrix{ \zeta_-\cr \zeta_+ }\right) = \left(\matrix{g^{1 \over 4}\, \lambdaa\,  \b s_-(\b z) \cr   \lambdaa^{-1} \, }\right)~,
}
where we have set $s_+=1$ and $s_-$ is a section of $\CO(2)$. Choosing $s_-=z$, which has zeros at the poles, we have a non-singular background for any metric satisfying the boundary conditions \metricnearzero. A similar story holds for $n=-m=-1$.

\newsec{Backgrounds Preserving Two Supercharges}

In this section, we study backgrounds which allow for a second Killing spinor $\t\zeta_\pm$ solving \KSEzetat. In the presence of $\zeta_\pm$ and $\t\zeta_\pm$ one can construct a complex vector of zero $R$-charge
\eqn\defK{K^\mu= \zeta\gamma^\mu\t\zeta~.}
 In the frame basis, $K_1 = \zeta_-\t\zeta_-$ and $K_{\b 1}= -\zeta_+\t\zeta_+$.
It follows from \KSEzeta, \KSEzetat\ that $K$ is Killing. We have several possibilities:
\medskip
\item{$\bullet$} If $K=0$, we have the ordinary topological $A$- or $\b A$-twist. This is what happens on Riemann surfaces of genus $\gen>1$, which do not admit Killing vectors.
\item{$\bullet$} If $K$ and $\b K$ are linearly independent and $[K, \b K]=0$, we are on the flat torus.
\item{$\bullet$} If $K$ and $\b K$ are linearly independent and $[K, \b K]\neq 0$, we are on the sphere with its round metric, and we actually have four supercharges \refs{\BeniniUI,\DoroudXW}. This case will be discussed in section 5.  
\item{$\bullet$} If $K$ and $\b K$ are linearly dependent, we are either on the sphere or on the torus with a $U(1)$ isometry. This is the most interesting case. (We focus on the sphere in the following, the torus is comparatively trivial.)
\medskip
\noindent  Consider the Riemann sphere. Similarly to the previous subsection, we can restrict ourselves to the ansatz
\eqn\partialsolii{
\t\zeta_-(z, \b z) = {1\over \lambdav(z,\bz) \lambdaa(z,\bz)} g^{1-n \over 8}\, \b t_-(\b z)~, \qquad 
\t\zeta_+(z, \b z) = {\lambdaa(z,\bz) \over \lambdav(z,\bz)} g^{1+n \over 8} \, t_+(z)
}
for the Killing spinor $\t\zeta$, where $t_\pm$ are holomorphic sections of the following line bundles:
\eqn\sectionsfCPii{
t_- \in \Gamma\left(\CO(-n+1) \right)~, \qquad t_+\in \Gamma\left(\CO(n+1) \right)~.
}
The Killing vector \defK\ is then 
\eqn\Kagain{K = -2 s_+(z)t_+(z) ~\d_z + 2 \b s_-(\bz) \b t_-(\bz) ~\d_\bz~.}
Without loss of generality, let us consider the real coordinates $\theta, \varphi$ on the sphere,  $\theta \in [0,\pi]$, $\varphi \sim \varphi + 2\pi$, with complex coordinate
\eqn\choosez{
z= h(\theta)\, e^{i\varphi}~.
}
Here $h(\theta)$ is a smooth real positive function with the same asymptotics as  $\tan{{\theta\over 2}}$ at $\theta=0,\pi$. By assumption, $K$ lies along the azimuthal direction $\d_\varphi = i (z\d_z -\bz \d_\bz)$.

Solving for the background fields in terms of $\t\zeta_\pm$, we have the same $A_\mu$ as in \backsol\ while $\CH, \t\CH$ are given by 
\eqn\backsolii{
\CH = {2\over \sqrt g} \d_z \left( g^{1+ n\over 4}{ \lambdaa^2 \, t_+ \over \b t_-}\right)~,
\qquad \qquad  \t\CH={2\over \sqrt g} \d_\bz \left( g^{1-n \over 4} {\b t_-\over \lambdaa^2\, t_+}\right)~.
}
In order for the background to preserve two supercharges of opposite $R$-charge,
\backsolii\ must be compatible with \backsol. 
If we further impose that $\lambdaa$ be invariant along $K$, \backsolii\ is compatible with \backsol\ if and only if
\eqn\compatcond{
\b s_- \b t_- = c_0\, \bz~, \qquad s_+ t_+ =c_0\, z~, \qquad s_+\, \d_z t_+ = \b t_-\, \d_\bz \b s_-~, 
\qquad  t_+\, \d_z s_+ = \b s_-\, \d_\bz \b t_-~,
}
with $c_0$ a complex constant. Note that $s_+ t_+$ and $s_- t_-$ both correspond to the holomorphic section of the line bundle $\CO(2)$ with zeros at the north and south poles ($\theta=0, \pi$ respectively). One easily checks that any regular solution of the first two equations in \compatcond\  (for the allowed values of $n$, $n=0, \pm 1$) is also solution of the last two equations. We discuss such solutions more explicitly  in the following subsections.

\subsec{The $A$-Twist (Bis)}
Let us first address the case when $K=0$ in \defK. This can only happen for $\zeta_-=0$ or $\zeta_+=0$, corresponding to the $A$- or the $\b A$-twist \backgrfieldsAtwist, \backgrfieldsAbartwist, which are defined on any Riemann surface $\Sigma_\gen$. The topological $A$-twist corresponds to
\eqn\sectionsAtwist{s_- =0~,\qquad s_+ = 1~, \qquad t_-=1~, \qquad t_+=0~,}
in the ansatz \partialsol, \partialsolii\ with $n=1$. Here $s_+$ and $t_-$ are sections of the trivial line bundle. Note that given the background \backgrfieldsAtwist\ we automatically have a second Killing spinor $\t\zeta_-\propto{1\over \zeta_+}$, $\t\zeta_+=0$. The case of the $\b A$-twist is similar. In these backgrounds, the $R$-charge is quantized in units of ${1\over {\gen-1}}$ due to \Rquantized.

Note that for the $A$-twist we have the freedom of turning on an arbitrary $\t\CH$, and similarly with an arbitrary $\CH$ for the $\b A$-twist. This is a rather mild deformation. 
In particular, one can show that the partition function does not depend on any continuous deformations of $\t\CH$ because such deformations are $Q$-exact in this case. The supersymmetry algebra is also unaffected.  
However, if there is a non-trivial flux for $\t\CH$ the quantization condition \quantZ\ holds and constrains the central charge.

\subsec{The  $\Omega$-Background on the Sphere}
On the sphere, the $A$-twist admits an interesting  $U(1)$-equivariant deformation, corresponding to turning on $s_-$ and $t_+$ in \sectionsAtwist, which are holomorphic sections of the holomorphic tangent bundle $T\C\P^1\cong \CO(2)$. 
Let us consider the azimuthal Killing vector 
\eqn\defV{
V= \d_\varphi= i \left(z\d_z -\bz\d_\bz \right)~.
}
The Killing vector \defK\ is proportional to \defV,
\eqn\KpropV{
K= \epsdef  V~, \qquad  \epsdef\in \C~,}
and equation \compatcond\  implies 
\eqn\sectionsAtwistdef{s_- = i {\b \epsdef\over 2}z~,\qquad s_+ = 1~, \qquad t_-=1~, \qquad t_+= -i { \epsdef\over 2} z~.}
For $\epsdef=0$ we recover the ordinary $A$-twist \sectionsAtwist.
The Killing spinors \partialsol, \partialsolii\ can be conveniently written in terms of  \defV,
\eqn\KSEquivAtwist{
\zeta = \left(\matrix{\zeta_- \cr \zeta_+}\right)=  \left(\matrix{\epsdef V_1 \cr 1}\right)~,     \qquad\qquad  
\t\zeta= \left(\matrix{\t\zeta_- \cr \t\zeta_+}\right)= \left(\matrix{1 \cr -\epsdef V_{\b 1}}\right)~.
}
(Here we have set $\lambdav=\lambdaa=1$ for simplicity.)
The corresponding background supergravity fields are
\eqn\bkgEquivAtwist{
 A_\mu = {1\over 2} \omega_\mu\,,  \qquad\quad \CH =  -i{\epsdef\over 2}\, \epsilon^{\mu\nu}\d_\mu V_\nu~, \qquad\quad \t\CH=0~.
}
The background $R$-symmetry gauge field takes its standard $A$-twist value with $m=-1$ unit of flux through the sphere (in our conventions). Therefore the $R$-charge must be integer quantized.  The equivariant deformation $\epsdef\neq 0$ corresponds to a non-trivial expectation value for the graviphoton $C_\mu$,
\eqn\graviphOmega{
C_\mu= {\epsdef\over 2} V_\mu~,  \qquad\quad  \t C_\mu = 0~,
}
up to a gauge transformation.   Note that $\CH$ has vanishing flux through the sphere, therefore the central charge is not constrained in this background.

It is instructive to consider the supersymmetry algebra corresponding to \KSEquivAtwist. The $A$-twist has the effect of twisting the spin  $S$ to $S'=S+{R\over 2}$, where $R$ is the $R$-charge. In particular the twisted spins of the supersymmetry parameters $\zeta_-$, $\zeta_+$, $\t\zeta_-$, $\t\zeta_+$ are  $1$, $0$, $0$, $-1$, respectively. 
On a field $\varphi_{(r,z,\t z,s)}$ of $R$-, $Z$-, $\t Z$-charges $r, z, \t z$ and spin $s$,  we have the supersymmetry algebra
\eqn\SUSYalgEquivAtwist{\eqalign{
&\{ \delta_\zeta, \delta_{\t \zeta}\} \varphi_{(r,z,\bz,s)} = - 2i   \big( z + \epsdef\CL_{V}|_{s\to s'}  \big) \varphi_{(r,z,\bz,s)} ~,\cr
&\delta_\zeta^2\, \varphi_{(r,z,\bz,s)} = 0~, \qquad\qquad\qquad\delta_{\t \zeta}^2\,\varphi_{(r,z,\bz,s)}= 0~,
}}
where $\CL_{V}|_{s\to s'}$ is the Lie derivative along $V$ with the spin $s$ replaced by the twisted spin $s'=s+{r\over 2}$. (See Appendix A for the definition of the Lie derivative.) This result easily follows by plugging the Killing spinors \KSEquivAtwist\ and background fields \bkgEquivAtwist-\graviphOmega\ in the more general supersymmetry algebra 
(6.1) to be introduced below. It is manifest from \SUSYalgEquivAtwist\ that the background \bkgEquivAtwist-\graviphOmega\ realizes a $U(1)$-equivariant deformation of the $A$-twist on $S^2$ with equivariant parameter $\epsdef$. This is also known as the $\Omega$-background ---see for instance \refs{\NekrasovRJ,\ShadchinYZ,\GerasimovEB, \HellermanMV}.

\subsec{The Sphere Without $R$-symmetry Flux}
Another background of recent interest is the sphere without flux for the $R$-symmetry gauge field. In this case, $s_\pm, t_\pm$ are sections of $\CO(1)$, and the most general solution  to \compatcond\  is either 
\eqn\sectionsnzero{s_- =1~,\qquad s_+ = i z~, \qquad t_-= - i z~, \qquad t_+= 1~.}
or  
\eqn\sectionsnzeroii{s_- =-i z~,\qquad s_+ = 1~, \qquad t_-= 1~, \qquad t_+= i z~.}
(The factors of $i$ are chosen for future convenience.)
Let us consider \sectionsnzero\ for definiteness. The  Killing spinors  \partialsol, \partialsolii\ read
\eqn\KSzerofluxsphere{
\zeta = \left(\matrix{\zeta_- \cr \zeta_+}\right)= g^{1\over 8} \left(\matrix{ \lambdaa \cr i \lambdaa^{-1}  z}\right)~,     \qquad\qquad  
\t\zeta= \left(\matrix{\t\zeta_- \cr \t\zeta_+}\right)= g^{1\over 8} \left(\matrix{ i \lambdaa^{-1} \b z \cr  \lambdaa}\right)~,
}
where we set $\lambdav=1$ for simplicity, and  the Killing vector \defK\ is $K = -2 \d_\varphi$. The corresponding supergravity fields can be obtained from \backgrfieldsgen, and one can check that $\CH, \t\CH$ are regular at the poles. Therefore we have two supercharges on any squashed sphere with $U(1)$ isometry and vanishing $R$-flux. 

As a concrete example of such a supersymmetric squashed sphere, consider the metric
\eqn\metricGL{
ds^2  = f(\theta)^2 d\theta^2 + \sin^2\theta d\varphi^2 = c(z, \bz)^2 dz d\bz~,
} 
with $f(\theta)$ an arbitrary smooth function such that $f\sim 1 + o(\theta^2)$ at $\theta=0$, and similarly at $\theta=\pi$.  We must have
\eqn\cgetc{
c(z,\bz) = g^{1\over 4}= {\sin\theta\over h(\theta)}~, \qquad f(\theta) = \sin\theta\,  {h'(\theta)\over h(\theta)}~,
}
with $h(\theta)$ the function introduced in \choosez. 
A convenient choice of $\lambdaa$ in \KSzerofluxsphere\ is
\eqn\choicelambdaa{
\lambdaa = \sqrt{{h(\theta)\over \tan{\theta\over 2}}}~.
}
In this case, the background supergravity fields take the simple form
\eqn\label{
A_\mu dx^\mu = \half \left(1 - {1\over f(\theta)}\right) d\varphi~, \qquad \CH= \t\CH = {i\over f(\theta)}~.
}
This background was studied in \GomisWY. Note that $\int_{S^2} dA=0$ and that we could set $A_\mu=0$ by choosing $\lambdaa$ to be constant. On the other hand we have non-trivial fluxes for the graviphotons, leading to the quantization condition
\eqn\quantZStwo{R_{S^2} \Im (z- \t z) \in \Z}
for the central charge. Here we restored the overall radius $R_{S^2}$ in the metric \metricGL\ for dimensional reasons.

\newsec{Maximally Supersymmetric Backgrounds}
It follows from \KSEzeta, \KSEzetat\ that a supersymmetric background preserves four supercharges (two $\zeta$'s and two $\t \zeta$'s) if and only if
\eqn\intcondtwozeta{ 
\d_\mu A_\nu - \d_\nu A_\mu=0~, \qquad \half R = \CH\t\CH~, \qquad
 \del_\mu \CH=0~, \qquad \del_\mu \t\CH=0~,
}
with $R$ the Ricci scalar.
In particular, the two-manifold must have constant scalar curvature. For compact two-manifolds, \intcondtwozeta\ can only be satisfied on the round sphere of radius $R_{S^2}$ with $\CH\t\CH = -{1\over R_{S^2}^2}$, or on the flat torus with $\CH\t\CH=0$.

Up to a $U(1)_R$ gauge transformation, the most general maximally supersymmetric $S^2$ background is
\eqn\spherefourQ{
ds^2 = {4 R_{S^2}^2 \over (1 + |z|^2)^2} dz d\bz~, \qquad A_\mu =0~, \qquad \CH = {i\over R_{S^2}}\lambdaa^2~, \qquad \t\CH= {i\over R_{S^2}} \lambdaa^{-2}~,
}
with $\lambdaa\in \C$ an arbitrary constant. The four Killing spinors correspond to the two choices of sections \sectionsnzero\ and \sectionsnzeroii, which are mutually compatible on the background \spherefourQ. In terms of the real coordinates $\theta,\varphi$ with $z= \tan{\theta\over 2}~e^{i\varphi}$, they read, up to a constant normalization, 
\eqn\fourKSstwo{\eqalign{
&\zeta =  \left(\matrix{\lambdaa  \cos{\theta\over 2}  \cr i \lambdaa^{-1} \sin{\theta\over 2}e^{i\varphi} }\right)~,\qquad\; \eta =  \left(\matrix{ i \lambdaa \sin{\theta\over 2}e^{-i\varphi} \cr   \lambdaa^{-1} \cos{\theta\over 2}}\right)~, \cr
&\t\zeta = \left(\matrix{i\lambdaa^{-1} \sin{\theta\over 2}e^{-i\varphi}  \cr \lambdaa \cos{\theta\over 2} }\right)~,\qquad \t\eta = \left(\matrix{\lambdaa^{-1} \cos{\theta\over 2}  \cr  i\lambdaa \sin{\theta\over 2}e^{i\varphi}}\right)~.
}}
The three Killing vectors  $\zeta \gamma^\mu \t\eta$, $\eta \gamma^\mu \t\zeta$ and $\zeta \gamma^\mu \t\zeta =\eta \gamma^\mu \t\eta$ generate the $SO(3)$ isometry of the round sphere.

Relaxing the requirement that the two-manifold is compact, an interesting one-complex-parameter family of backgrounds with four supercharges is provided by the $\Omega$-background on $\R^2$, which was studied from the field theory viewpoint in \refs{\ShadchinYZ}. This background can be obtained as the flat space limit of the $\Omega$-background on $S^2$ of section 4.2, focusing on the north pole patch.  The metric is flat and all the other supergravity background fields vanish except for the graviphoton \graviphOmega, which leads to a constant background for $\CH=-i\epsdef$. In addition to the two supercharges \KSEquivAtwist\ that exist on curved space, the flat space $\Omega$-background preserves two ordinary flat space supercharges. The four Killing spinors are: 
\eqn\fourKSstwo{\eqalign{
&\zeta = \left(\matrix{\epsdef V_z \cr 1}\right)~,\qquad\; \eta = \left(\matrix{1 \cr 0}\right)~,\cr
&\t\zeta = \left(\matrix{1 \cr -\epsdef V_{\bz}}\right)~,\qquad \t\eta = \left(\matrix{0 \cr 1}\right)~.
}}
From those Killing spinors we can build the Killing vectors
\eqn\KvecOmega{
\zeta \gamma^\mu \t\zeta\, \d_\mu = \epsdef V~, 
\qquad \zeta \gamma^\mu \t\eta\, \d_\mu= - 2 \d_z~,\qquad
\eta \gamma^\mu \t\zeta\, \d_\mu= 2\d_\bz~,  }
with $V$ the rotational Killing vector \defV. Note that with our present definition of the $\Omega$-background, we obtain supersymmetric actions which differ from the ones of {\it e.g.} \refs{\ShadchinYZ, \DimofteTZ}.%
\foot{The main difference is that we effectively shift the central charge $z$ to $z + \epsdef \CL_V$ while keeping $\t z$ fixed, which is allowed in Euclidean signature, while in \refs{\ShadchinYZ, \DimofteTZ} both $z$ and 
$\t z$ 
are being shifted.} It would be interesting to understand better the relation between the two approaches.

\newsec{Supersymmetry Multiplets and Supersymmetric Lagrangians}
In this section, we  study the curved-space generalization of the standard $\CN=(2,2)$ supersymmetry multiplets. Recall that two-dimensional Lorentz invariance (or rotation invariance, in our case) allows for more general supersymmetry multiplets than in higher dimensional theories with the same amount of supersymmetry \refs{\GatesNK,\BuscherUW}.
We will discuss in detail the chiral  and twisted chiral multiplets, possibly coupled to vector or twisted vector multiplets, respectively. Note that a symmetry that acts only on chiral multiplets can be gauged with an ordinary vector multiplet, while a symmetry that acts only on the twisted chiral multiplets can be gauged with a twisted vector multiplet. The gauging of more general symmetries can also be considered, but it requires more complicated vector multiplets \refs{\LindstromVC,\LindstromHX} which we will not consider in this work. Yet another family of interesting $\CN=(2,2)$ multiplets are the semi-chiral multiplets \BuscherUW, which we briefly discuss in Appendix E.

\subsec{Supersymmetry Algebra}
The generalization of the flat space  $\CN=(2,2)$ supersymmetry algebra \susyflat\ to any curved-space supersymmetric background of the kind discussed in previous sections is:
\eqn\rigidsalg{\eqalign{
& \{ \delta_\zeta, \delta_{\t \zeta}\} \varphi_{(\fr,\fz,\t\fz)} = - 2i \left[ \CL'_{K}  + \zeta_+\t\zeta_- \zH - \zeta_-\t\zeta_+ \zHt   \right]\varphi_{(\fr,\fz,\t\fz)}~,\cr
& \{ \delta_\zeta, \delta_\eta\} \varphi_{(\fr,\fz,\t\fz)} = 0~, \qquad \{ \delta_{\t \zeta}, \delta_{\t \eta}\} \varphi_{(\fr,\fz,\t\fz)}= 0~.}}
Here~$\varphi_{(\fr,\fz,\t\fz)}$ is a field of arbitrary spin, $R$-charge~$\fr$, and central charge~$\fz, \t\fz$. We use~$\CL'_K$ to denote a modified Lie derivative along~$K^\mu$, which is covariant under local~$R$-, $Z$-and $\t Z$-transformations,
\eqn\hatliedef{ \CL'_K \varphi_{(\fr,\fz,\t\fz)} =\left( \CL_K - i \fr K^\mu A_\mu   +\half \fz K^\mu \t C_\mu  -\half \t\fz K^\mu C_\mu \right)\varphi_{(\fr,\fz,\t\fz)}~,}
with $ K^\mu = \zeta \gamma^\mu \t \zeta$ the Killing vector \defK.
The most straightforward way to derive \rigidsalg\ is by twisted dimensional reduction of the three-dimensional algebra of \ClossetRU, as we explain in Appendix C.

\subsec{General Multiplet}
Let us consider a general complex multiplet $\CS$ whose bottom component is a complex scalar or $R$-charge $\fr$  and central charges $\fz$, $\t\fz$, with components
\eqn\Scomponents{
\CS = \big(C,\chi_\pm, \t\chi_\pm,  M, \t M, a_\mu, \sigma,\t\sigma, \lambda_\pm, \t\lambda_\pm, D\big)~.
}
It has $8+8$ complex components.
In flat space it is represented by the unconstrained complex superfield 
\eqn\CplxSuperf{\eqalign{
\CS =
&\; C +i \tm \chi_- + i \tp \chi_+ + i\ttm \t\chi_- + i\ttp \t\chi_+ + i\tp\tm M+i\ttp\ttm \t M \cr
&-2(\tm\ttm a_z-\tp\ttp a_\bz)+i\tm\ttp\sigma-i\tp\ttm\t\sigma+2i\ttp\ttm\tm(\lambda_-+i\d_z\t\chi_+)\cr &+2i\ttp\ttm\tp(\lambda_+ + i\d_\bz\t\chi_-) -2i\tp\tm\ttm (\t\lambda_--i\d_z\chi_+) \cr 
&-2i\tp\tm\ttp (\t\lambda_+ -i\d_\bz\chi_-) -2\tp\tm\ttp\ttm (D+2\d_z\d_\bz C)~.
}}
In the following, we spell out its curved space generalization. We work in the complex frame $e^1, e^{\b 1}$ (see Appendix A) and express all tensors, including covariant derivatives, in the frame basis.
Let us also define the $R$- and $Z, \t Z$-covariant derivative 
\eqn\defDfullcov{
D_\mu \varphi_{(r,z,\t z)} = \left(\nabla_\mu - i r A_\mu  + \half z \t C_\mu - \half \t z C_\mu\right)\varphi_{(r,z,\t z)}~,
}
acting on any field $\varphi_{(r,z,\t z)}$ of $R$-charge $r$ and complex central charge $z, \t z$.%
\foot{In this section $z$ always denotes the central charge and not a complex coordinate. Since we are working in the frame basis this should cause no confusion.} 
The curved space supersymmetry algebra is represented on the general multiplet \Scomponents\ by: 
\eqn\SusyCplxSuperf{\eqalign{
\delta C &= i (\zeta_+\chi_- - \zeta_-\chi_+)+i (\t\zeta_+\t\chi_- - \t\zeta_-\t\chi_+)  \, ,\cr
\delta \chi_- &= \zeta_- M -\t\zeta_- \left(\sigma + \left(z-{r\over 2} \CH\right) C \right)  + 2\t\zeta_+ \left(D_1 C+ia_1\right) \, ,\cr
\delta \chi_+ &= \zeta_+ M -\t\zeta_+ \left(\t \sigma + \left(\t z-{r\over 2} \t\CH\right) C \right)  + 2\t\zeta_- \left(D_{\b 1} C+ia_{\b 1}\right)\, ,\cr
\delta \t\chi_- &= \t\zeta_- \t M -\zeta_- \left(\t\sigma  - \left(\t z-{r\over 2} \t\CH\right) C \right)  + 2\zeta_+ \left(D_1 C- i a_1\right)\, , \cr
\delta \t\chi_+ &= \t\zeta_+ \t M -\zeta_+ \left( \sigma - \left( z-{r\over 2} \CH\right) C \right)  + 2\zeta_- \left(D_{\b 1} C-ia_{\b 1}\right)\, ,\cr
\delta M &= -2(\t\zeta_+\t\lambda_-- \t\zeta_-\t\lambda_+) + 2 i   \zHt \t\zeta_+ \chi_- - 2i \zH \t\zeta_- \chi_+ 
      \cr  &\;\quad  + 4 i \t\zeta_+ D_1 \chi_+ - 4 i \t\zeta_- D_{\b 1} \chi_-\, ,\cr
\delta \t M &= 2(\zeta_+\lambda_- - \zeta_-\lambda_+) - 2 i   \zH \zeta_+ \t\chi_- + 2i \zHt \zeta_- \t\chi_+ 
      \cr  &\;\quad  + 4 i \zeta_+ D_1 \t\chi_+ - 4 i \zeta_- D_{\b 1} \t\chi_-\, ,
}}
\eqn\SusyCplxSuperfii{\eqalign{
\delta a_1 &= - i  \zeta_- \t\lambda_- -i \t\zeta_- \lambda_- + D_1 \left( \zeta_+\chi_- - \zeta_-\chi_+ - \t\zeta_+\t\chi_- + \t\zeta_-\t\chi_+\right) \, ,   \cr
\delta a_{\b 1} &=  i  \zeta_+ \t\lambda_+ +i \t\zeta_+ \lambda_+ + D_{\b 1} \left( \zeta_+\chi_- - \zeta_-\chi_+ - \t\zeta_+\t\chi_- + \t\zeta_-\t\chi_+\right) \, ,   \cr
\delta \sigma &=  2\zeta_- \t\lambda_+ + 2\t\zeta_+ \lambda_- + i \zH \left( \zeta_+\chi_- - \zeta_-\chi_+ - \t\zeta_+\t\chi_- + \t\zeta_-\t\chi_+\right)  \, , \cr
\delta \t\sigma &=  - 2\t\zeta_- \lambda_+ - 2\zeta_+ \t\lambda_- + i \zHt \left( \zeta_+\chi_- - \zeta_-\chi_+ - \t\zeta_+\t\chi_- + \t\zeta_-\t\chi_+\right)  \, , \cr
\delta \lambda_- &= i \zeta_- \left(  D - 2i (D_{\b 1}a _1 - D_1 a_{\b 1}) + \t\CH \sigma +\half\zHt\sigma -\half \zH \t\sigma \right)  
\cr & \;\quad  + 2 i \zeta_+ \left(D_1 \sigma - i \zH a_1  \right) \, , \cr
\delta \lambda_+ &= i \zeta_+ \left(  D + 2i (D_{\b 1}a _1 - D_1 a_{\b 1}) + \CH \t\sigma -\half\zHt\sigma +\half \zH \t\sigma \right)  
\cr & \;\quad  + 2 i \zeta_- \left(D_{\b 1} \t\sigma - i \zHt a_{\b 1}  \right) \, , \cr
\delta \t\lambda_- &= -i \t\zeta_- \left(  D + 2i (D_{\b 1}a _1 - D_1 a_{\b 1}) + \CH \t\sigma +\half\zHt\sigma -\half \zH \t\sigma \right)  
\cr & \;\quad  - 2 i \t\zeta_+ \left(D_1 \t\sigma - i \zHt a_1  \right) \, , \cr
\delta \t\lambda_+ &= -i \t\zeta_+ \left(  D - 2i (D_{\b 1}a _1 - D_1 a_{\b 1}) + \t\CH \sigma -\half\zHt\sigma +\half \zH \t\sigma \right)  
\cr & \;\quad  - 2 i \t\zeta_- \left(D_{\b 1} \sigma - i \zH a_{\b 1}  \right) \, , \cr
\delta D &= - 2 D_1\left( \zeta_+ \t\lambda_+ - \t\zeta_+ \lambda_+ \right) + 2 D_{\b 1}\left( \zeta_- \t\lambda_- - \t\zeta_- \lambda_- \right) \cr
           &\quad  + \zH \left( \zeta_+ \t\lambda_- - \t\zeta_- \lambda_+ \right)  -\zHt \left( \zeta_- \t\lambda_+ - \t\zeta_+ \lambda_- \right) \cr
&\quad + i \left( {r\over 4} R -\half \CH \t z -\half \t\CH z\right)\left( \zeta_+\chi_- - \zeta_-\chi_+ - \t\zeta_+\t\chi_- + \t\zeta_-\t\chi_+\right) \, .
}}
In the last line, $R$ stands for the Ricci scalar of the two-manifold.  These transformations realize the algebra \rigidsalg\ for any spinors $\zeta_\pm$, $\t\zeta_\pm$ satisfying the Killing spinor equations.

\subsec{Chiral and Twisted Chiral Multiplets}
A chiral multiplet $\Phi$ of $R$-charge $r$ and central charges $z$, $\t z$ is a general multiplet satisfying the constraints
\eqn\defChiral{\t\chi_-=\t\chi_+ =0~.}
This is equivalent to the superspace constraint $\t D_\pm \Phi =0$ on the corresponding superfield in flat space.
The chiral multiplet has components
\eqn\compoPhi{
\Phi= \big(\phi, \psi_-, \psi_+, F\big)~,
}
of $R$-charges $r, r-1, r-1, r-2$, respectively, whose embedding into the general multiplet \Scomponents\ is given in Appendix D.
Its supersymmetry transformations are
\eqn\ChiralSusy{\eqalign{
\delta \phi &= \sqrt{2} (\zeta_+\psi_- -\zeta_-\psi_+)~,\cr
\delta \psi_- &= \sqrt{2} \zeta_- F - i\sqrt{2}  \t\zeta_- \zH \phi +2i \sqrt2 \t\zeta_+ D_1\phi~,\cr
\delta \psi_+ &= \sqrt{2} \zeta_+ F - i\sqrt{2}   \t\zeta_+ \zHt \phi +2i \sqrt2 \t\zeta_- D_{\b 1}\phi~,\cr
\delta F &= \sqrt2 i  \left(\t z-{r-2\over 2} \t\CH\right) \t\zeta_+\psi_- - \sqrt2 i  \left( z-{r-2\over 2} \CH\right)  \t\zeta_-\psi_+ \cr
& \quad +2i \sqrt2 D_1(\t\zeta_+ \psi_+) - 2i \sqrt2 D_{\b 1}(\t\zeta_-\psi_-)~.
}}
Similarly, an antichiral multiplet $\t\Phi$ of $R$-charge $-r$ and central charge $-z$, $-\t z$ is a general multiplet satisfying  the constraints
\eqn\defAntiChiral{\chi_-=\chi_+ =0~,}
or $D_\pm \t\Phi=0$ in flat space. It has components
\eqn\compoPhitilde{
\t\Phi= \big(\t\phi, \t\psi_-, \t\psi_+, \t F\big)~,
}
of $R$-charges $-r, -r+1, -r+1, -r+2$, respectively.
 Its embedding into \Scomponents\ is given in Appendix D and its supersymmetry transformations are 
\eqn\AntiChiralSusy{\eqalign{
\delta \t\phi &= -\sqrt{2} (\t\zeta_+\t\psi_- -\t\zeta_-\t\psi_+)~,\cr
\delta \t\psi_- &= \sqrt{2} \t\zeta_- \t F + i\sqrt{2} \zeta_- \zHt \t\phi -2i \sqrt2 \zeta_+ D_1\t\phi~,\cr
\delta \t\psi_+ &= \sqrt{2} \t\zeta_+ \t F + i\sqrt{2} \zeta_+ \zH \t\phi -2i \sqrt2 \zeta_- D_{\b 1}\t\phi~,\cr
\delta \t F &= \sqrt2 i  \left(z-{r-2\over 2}\CH\right)\zeta_+\t\psi_- - \sqrt2 i \left( \t z-{r-2\over 2} \t\CH\right) \zeta_-\t\psi_+ \cr
& \quad +2i \sqrt2 D_1(\zeta_+ \t\psi_+) - 2i \sqrt2 D_{\b 1}(\zeta_-\t\psi_-)~.
}}

One can also define the twisted chiral multiplet $\Omega$, which is a general multiplet satisfying
\eqn\defTwistedChiral{\chi_-= \t\chi_+=0~,}
or $D_-\Omega = \t D_+ \Omega=0$ in flat space. Note that this multiplet is special to two dimensions.
It turns out that such a multiplet can only be embedded into a general multiplet \Scomponents\ of vanishing $R$- and $Z, \t Z$-charges, $r=z=\t z=0$. It has components 
\eqn\compoOmega{
\Omega=\big(\omega, \eta_- , \t\eta_+, G\big)~,
}
of $R$-charges $0, 1, -1,0$, respectively. Its embedding into \Scomponents\ is given in Appendix D and its
 supersymmetry variations are 
\eqn\TwistedChiralSusy{\eqalign{
\delta \omega &= \sqrt{2} (\t\zeta_+\eta_- - \zeta_-\t\eta_+)~, \cr
\delta \eta_- &= \sqrt{2} \zeta_- G +2i\sqrt{2}  \zeta_+ D_1\omega~, \cr
\delta \t\eta_+ &= \sqrt{2} \t\zeta_+ G +2i \sqrt{2}  \t\zeta_-D_{\b 1}\omega~, \cr
\delta G &= 2i \sqrt{2}  (\zeta_+ D_1 \t\eta_+ - \t\zeta_- D_{\b 1} \eta_-)\,.
}}
The twisted antichiral multiplet  $\t \Omega$ is similarly defined by
\eqn\defTwistedChiral{\chi_-= \t\chi_+=0~,}
or  $D_-\t\Omega= \t D_+ \t\Omega=0$ in flat space. It has components
\eqn\compoOmega{
\t\Omega=  \big(\t\omega, \t\eta_- , \eta_+, \t G\big)~,
}
of $R$-charges $0, -1, 1, 0$, respectively.
Its embedding into \Scomponents\ is given in Appendix D and its supersymmetry transformations are 
\eqn\TwistedAntiChiralSusy{\eqalign{
\delta \t\omega &= -\sqrt{2} (\zeta_+\t\eta_- - \t\zeta_-\eta_+)~, \cr
\delta \t\eta_- &= \sqrt{2} \t\zeta_- \t G -2i\sqrt{2}  \t\zeta_+ D_1\t\omega~, \cr
\delta \eta_+ &= \sqrt{2} \zeta_+ \t G -2i \sqrt{2}  \zeta_-D_{\b 1}\t\omega~, \cr
\delta \t G &= 2i \sqrt{2}  (\t\zeta_+ D_1 \eta_+ - \zeta_- D_{\b 1} \t\eta_-)\,.
}}

The chiral and twisted chiral multiplet both have $2+2$ complex components. 
Another interesting set of constrained multiplets are given by the semi-chiral multiplets, which are general multiplets with either $\chi_+, \chi_-, \t\chi_-$ or $\t\chi_+$ set to zero. They have $4+4$ complex components. We briefly discuss them in Appendix  E. Finally, we could also consider a stronger constraint which sets three of the $\chi_\pm, \t\chi_\pm$ to zero. For instance, consider the case $\t\chi_-=\t\chi_+ = \chi_+=0$. Such an ultra-short multiplet  has $1+1$ components $\left(\varphi, \psi_-\right)$ of vanishing $R, Z, \t Z$-charges which are constrained to be holomorphic:
\eqn\holoconst{\d_{\b 1} \varphi =D_{\b 1} \psi_-=0~.}
They have the supersymmetry transformations
\eqn\susychiralbos{
\delta \varphi = \sqrt{2} \zeta_+\psi_-~,\qquad
\delta \psi_- =2i \sqrt2 \, \t\zeta_+ \d_1\varphi~.
}
This multiplet is the supersymmetrization of the two-dimensional chiral boson (see for instance \FloreaniniAS). We will not discuss it any further in this work.

\subsec{Linear and Twisted Linear Multiplets}

The linear multiplet is a general multiplet $\CJ$ of vanishing $R$- and $Z, \t Z$-charges which satisfies the constraint \eqn\constrJ{M=\t M=0~.}
This corresponds to $D_+ D_- \CJ= \t D_+ \t D_- \CJ=0$ in flat space. 
It has components
\eqn\comCJ{
\CJ= (J, j_\pm, \t j_\pm, j_\mu, K, \t K)~,
}
of $R$-charges $0, -1, 1, 0, 0, 0$, where $j_\mu$ is a conserved current, $\nabla_1 j_{\b 1}+ \nabla_{\b 1} j_1=0$.
The embedding  of $\CJ$ into \Scomponents\ is given in Appendix D and its supersymmetry transformations are
\eqn\LinearMultSusy{\eqalign{
\delta J &= i (\zeta_+j_- - \zeta_-j_+) +  i (\t\zeta_+\t j_- - \t\zeta_-\t j_+) \, ,\cr
\delta j_- &= \t\zeta_- K + 2\t\zeta_+(D_1 J - i j_1)\, , \cr
\delta j_+ &= \t\zeta_+ \t K + 2\t\zeta_- (D_{\b 1} J - i j_{\b 1})\, , \cr
\delta \t j_- &= \zeta_- \t K + 2\zeta_+ (D_{1} J + i j_{ 1})\, , \cr
\delta \t  j_+ &= \zeta_+  K + 2 \zeta_- (D_{\b 1} J + i j_{\b 1})\, , \cr
\delta j_1 &= - D_1\left( \zeta_+j_- + \zeta_-j_+   -\t\zeta_+\t j_- - \t\zeta_-\t j_+ \right)\, , \cr
\delta j_{\b 1} &=  D_{\b 1} \left( \zeta_+j_- + \zeta_-j_+   -\t\zeta_+\t j_- - \t\zeta_-\t j_+ \right)\, , \cr
\delta K &= - 4 i \zeta_- D_{\b 1} j_- + 4 i \t\zeta_+ D_1 \t j_+ \, ,\cr
\delta \t K &=  -4 i \t\zeta_- D_{\b 1} \t j_- + 4 i \zeta_+ D_1  j_+ \,.
}}
Such a multiplet can be minimally coupled to the ordinary vector multiplet $\CV$ to be discussed in section 6.6 below.

The twisted linear multiplet $\h \CJ$ is a general multiplet of vanishing $R$- and $Z, \t Z$-charges which satisfies the constraint
 \eqn\constrhJ{\sigma=\t \sigma=0~,}
or $D_+\t D_- \h\CJ= D_- \t D_+ \h \CJ=0$ in flat space.
It has components
\eqn\comCJ{
\h \CJ= (\h J, \h j_\pm, \t {\h j}_\pm, \h K, \t {\h
 K}, \h j_\mu)~,
}
of $R$-charges $0, -1, 1, -2, 2, 0$, where $\h j_\mu$ is a conserved current, $\nabla_1 \h j_{\b 1}+ \nabla_{\b 1} \h j_1=0$.
The embedding  of $\h \CJ$ into \Scomponents\ is given in Appendix D and its supersymmetry transformations are
\eqn\LinearMultSusy{\eqalign{
\delta \h J &= i (\zeta_+ \h j_- - \zeta_- \h j_+) +  i (\t\zeta_+\t{\h j}_- - \t\zeta_-\t {\h j}_+) \, ,\cr
\delta \h j_- &= \zeta_- \h K + 2\t\zeta_+(D_1 \h J + i \h j_1)\, , \cr
\delta \h j_+ &= \zeta_+  \h K + 2\t\zeta_- (D_{\b 1} \h J - i \h j_{\b 1})\, , \cr
\delta \t {\h j}_- &= \t \zeta_- \t {\h K} + 2\zeta_+ (D_{1} \h J - i \h j_{ 1})\, , \cr
\delta \t {\h j}_+ &= \t \zeta_+ \t {\h K} + 2 \zeta_- (D_{\b 1} \h J + i \h j_{\b 1})\, , \cr
\delta \h j_1 &= D_1\left( \zeta_+ \h j_- - \zeta_- \h j_+   -\t\zeta_+\t{\h j}_- + \t\zeta_-\t {\h j}_+ \right)\, , \cr
\delta \h j_{\b 1} &= - D_{\b 1} \left( \zeta_+ \h j_- - \zeta_- \h j_+   -\t\zeta_+\t{\h j}_- + \t\zeta_-\t {\h j}_+ \right)\, ,\cr
\delta \h K &= -4 i \t\zeta_- D_{\b 1} \h j_- + 4 i \t\zeta_+ D_1 \h j_+ \, ,\cr
\delta \t {\h K} &=  -4 i \zeta_- D_{\b 1} \t{\h j}_- + 4 i \zeta_+ D_1 {\t {\h j}_+}~.
}}
Such a multiplet can be minimally coupled to the twisted vector multiplet $\h
\CV$ to be discussed in section 6.6 below.

\subsec{Multiplying Multiplets}
Starting from \SusyCplxSuperf-\SusyCplxSuperfii, it is straightforward to take the product of any two general multiplets. 
The product of two general multiplets   $\CS_1, \CS_2$ of lowest components $C_1, C_2$ of charges $r_1, z_1, \t z_1$ and $r_2, z_2, \t z_2$, respectively, is a general multiplet $\CS= \CS_1 \CS_2$ of charges $r_1 + r_2, z_1 + z_2, \t z_1 + \t z_2$ given by 
\eqn\productrulestwodim{\eqalign{
&C= C_1 C_2~, \cr
& \chi_\mp = \chi_{1 \mp} C_2 + C_1 \chi_{2\mp}~, \cr
& \t\chi_\mp = \t\chi_{1 \mp} C_2 + C_1 \t\chi_{2\mp}~, \cr
& M = M_1 C_2 + C_1 M_2 - i (\chi_{1+}\chi_{2 -}-\chi_{1-}\chi_{2 +})~,\cr
& \t M = \t M_1 C_2 + C_1 \t M_2 - i (\t\chi_{1+}\t\chi_{2 -}-\t\chi_{1-}\t\chi_{2 +})~,\cr
& a_1 = a_{11} C_2 + C_1 a_{21} -\half  (\chi_{1-}\t \chi_{2 -}-\t\chi_{1-}\chi_{2 -})~, \cr
& a_{\b 1} = a_{1\b 1} C_2 + C_1 a_{2\b 1} +\half  ( \chi_{1+}\t \chi_{2 +}-\t\chi_{1+}\chi_{2 +})~, \cr
&\sigma = \sigma_1 C_2 + C_1 \sigma_2 - i (\chi_{1-}\t \chi_{2 +}-\t\chi_{1+}\chi_{2 -})~,\cr
&\t\sigma = \t\sigma_1 C_2 + C_1 \t\sigma_2 + i (\chi_{1+}\t \chi_{2 -}-\t\chi_{1-}\chi_{2+})~,
}}
\eqn\productrulestwodimii{\eqalign{
& \lambda_- = \lambda_{1-} C_2 + {i\over 2} \t M_1 \chi_{2 -} + {i\over 2}\t\chi_{1 -} (\sigma_2 +  z_{2 \CH}C_2) - i \t\chi_{1+}(D_1 C_2 + i a_{2 1}) + (1\leftrightarrow 2)~, \cr
& \lambda_+ = \lambda_{1+} C_2 + {i\over 2} \t M_1 \chi_{2 +} + {i\over 2}\t\chi_{1 +} (\t\sigma_2 +\t z_{2 \CH}C_2) - i \t\chi_{1-}(D_{\b 1} C_2 + i a_{2\b 1}) + (1\leftrightarrow 2)~, \cr
& \t\lambda_- = \t\lambda_{1-} C_2 - {i\over 2}  M_1 \t\chi_{2 -} - {i\over 2}\chi_{1 -} (\t\sigma_2 -\t z_{2 \CH}C_2) + i \chi_{1+}(D_{1} C_2 - i a_{2 1}) + (1\leftrightarrow 2)~, \cr
& \t\lambda_+ = \t\lambda_{1+} C_2 - {i\over 2}  M_1 \t\chi_{2 +} - {i\over 2}\chi_{1 +} (\sigma_2 - z_{2 \CH}C_2) + i \chi_{1-}(D_{\b 1} C_2 - i a_{2 \b 1}) + (1\leftrightarrow 2)~,\cr 
&D = D_1 C_2 + C_1 D_2  + \half (\t M_1 M_2+ M_1 \t M_2) -\half (\t\sigma_1 \sigma_2 + \sigma_1 \t \sigma_2)  - 2 D_{\b 1 }C_1 D_1 C_2\cr
& \qquad - 2 D_1 C_1 D_{\b 1} C_2  +\half(\t z_{1\CH} z_{2\CH}+z_{1\CH}\t z_{2\CH})C_1 C_2  -2 (a_{1\b 1} a_{21}+ a_{2\b 1}a_{11})\cr
& \qquad - i D_{\b 1} \chi_{1-}\t\chi_{2-} + i \chi_{1-} D_{\b1} \t\chi_{2-} + i\t\chi_{1-}D_{\b 1}\chi_{2-} - i D_{\b 1}\t\chi_{1-}\chi_{2-}  \cr
& \qquad + i D_{1} \chi_{1+}\t\chi_{2+} - i \chi_{1+} D_{1} \t\chi_{2+} - i\t\chi_{1+}D_{ 1}\chi_{2+} + i D_{ 1}\t\chi_{1+}\chi_{2+}  \cr
& \qquad -{i\over 2}( z_{1\CH} -  z_{2\CH})(\chi_{1+}\t\chi_{2-}+ \t\chi_{1-}\chi_{2+})
           +{i\over 2}(\t z_{1\CH} - \t z_{2\CH})(\t\chi_{1+}\chi_{2-}+ \chi_{1-}\t\chi_{2+})\cr
&\qquad -(\lambda_{1+}\chi_{2-}- \lambda_{1-}\chi_{2+}) +(\chi_{1-}\lambda_{2+}-\chi_{1+}\lambda_{2-})\cr &\qquad+(\t\chi_{1+}\t\lambda_{2-}-\t\chi_{1-}\t\lambda_{2+}) - (\t\lambda_{1-}\t\chi_{2+}- \t\lambda_{1+}\t\chi_{2-}) ~,
}}
where we introduced the notation:
\eqn\defzh{
z_{i \CH}= \left( z_i - {r_i\over 2} \CH\right) ~, \qquad \t z_{i \CH}= \left(\t z_i - {r_i\over 2} \t
\CH\right)~, \qquad i=1,2~.
}

Let us discuss some important special cases.
The product of two chiral multiplets of $R$- and $Z, \t Z$-charge $(r_1, z_1, \t z_1)$ and $(r_2, z_2, \t z_2)$ give another chiral multiplet of charge $(r_1+r_2, z_1+z_2, \t z_1 + \t z_2)$. More generally, consider any holomorphic function $W(\phi^i)$, with $\phi^i$ the bottom components of some chiral multiplets. We have the associated chiral multiplet
\eqn\chiralmultW{
\left(\phi^W, \psi_-^W, \psi_+^W, F^W  \right) = \left(W,\,  \psi_-^i \d_i W, \,   \psi_+^i \d_i W , \, F^i \d_i W + \psi_-^i \psi_+^j \d_i \d_j W \right)\, ,
}
provided that $W$ is quasi-homogeneous with respect to the $R$- and $Z, \t Z$-symmetries, namely
\eqn\quasiHomConOnW{
\sum_i z_i\, \phi^i \d_i W  =\left(\sum_i z_i\right) W\, , \qquad \sum_i r_i \,\phi^i \d_i W  = \left(\sum_i r_i\right) W\,  .
}
We can also consider $\t W(\t\phi^i)$ an antiholomorphic function of antichiral multiplets.

Similarly, the product of two twisted chiral multiplets gives another twisted chiral multiplet. We can consider a general holomorphic function $\hat W(\omega^i)$, with $\omega^i$ the bottom components of some twisted chiral multiplets.  $\hat W$ is the bottom component of a twisted chiral multiplet with components
\eqn\TwistedchiralmultW{
\left(\omega^{\hat W}, \eta_-^{\hat W}, \t\eta_+^{\hat W}, G^{\hat W} \right) = \left( {\hat W},\,  \eta_-^i \d_i {\hat W},\, \t\eta_+^i \d_i {\hat W},\,   G^i \d_i {\hat W} +\eta_-^i \t\eta_+^j \d_i \d_j {\hat W}\right)\, ,
}
We can also consider $\t{\h W}(\t \omega^i)$ an antiholomorphic function of twisted antichiral multiplets. 
Note that the formulas \chiralmultW, \TwistedchiralmultW\ are like in flat space. (The only difference with flat space, which will be important below, is that the supersymmetry variation of the $G$-component of a twisted chiral multiplet is not a total derivative by itself.)

The product rules \productrulestwodim-\productrulestwodimii\ are easily generalized. In particular, given any number of general multiplets $\CS^a$ of  $R$- and $Z, \t Z$-charges $r^a, z^a, \t z^a$,
\eqn\Sacomponents{\CS^a = \big(C^a,\chi_\pm^a, \t\chi_\pm^a,  M^a, \t M^a, a_\mu^a, \sigma^a,\t\sigma^a, \lambda_\pm^a, \t\lambda_\pm^a, D^a \big)~,}
we can build a general multiplet $\CK(\CS^a)$ of lowest component $K= K(C^a)$ with $K$ any function.
Let us define
\eqn\defKabc{ K_{a_1 a_2 \cdots a_n} = {\d \over \d C^{a_1}}\cdots  {\d \over \d C^{a_n}} K~,}
which is totally symmetric in its indices.
We will consider the case of $K$ neutral and  quasi-homogeneous of degree zero:
\eqn\quasihom{
\sum_a r^a C^a K_a=0~, \quad  \sum_a z^a C^a K_a=0~, \quad \sum_a \t z^a C^a K_a=0~.
}
It is straightforward to extract the components of $\CK$. In particular, its $D$-term is given by
\eqn\DofK{\eqalign{
&D^\CK = \;  K_a \left( D^a - \half   z_{\CH}^a \t z_{\CH}^a C^a\right) \cr
&\qquad\quad+  K_{ab}\left( \half M^a \t M^b -\half \sigma^a \t\sigma^a - 2 D_1 C^a D_{\b 1} C^b - 2 a_{1}^a a_{\b 1}^b \right)~,\cr
&\quad\qquad+  K_{ab}\Big( i\chi_-^a D_{\b 1}\t\chi_-^b - i D_{\b 1} \chi_-^a\,\t\chi_-^b
- i\chi_+^a D_{1}\t\chi_+^b + i D_{ 1}\chi_+^a\,\t \chi_+^b \cr
&\qquad\qquad\qquad -{i\over 2} z_\CH^a \big(\t\chi_-^a \chi_+^b -\t\chi_-^b \chi_+^a \big) +{i\over 2} \t z_\CH^a \big(\chi_-^a \t\chi_+^b -\chi_-^b \t\chi_+^a \big)\cr
& \qquad \qquad\qquad  -\big(\lambda_+^a \chi_-^b - \lambda_-^a \chi_+^b \big)+\big(\t\lambda_+^a \t\chi_-^b - \t\lambda_-^a \t\chi_+^b \big)\Big)\cr
& \quad\qquad+\half K_{abc}\Big(  i \sigma^a \t\chi_-^b \chi_+^c + i \t\sigma^a \chi_-^b \t\chi_+^c + i M^a \t\chi_-^b \t\chi_+^c + i \t M^a \chi_-^b \chi_+^c \cr
& \quad\qquad\qquad\qquad + 2 a_1^a \t\chi_+ \chi_+^c - 2 a_{\b 1}^a \t\chi_-^b \chi_-^c\Big) 
\cr
& \quad\qquad -\half K_{abcd} \,\chi_+^a \chi_-^b \t\chi_+^c\t\chi_-^d~,
}}
where we sum over repeated indices (for instance, $K_a z_\CH^a \phi^a= \sum_a K_a z_\CH^a \phi^a$).

\subsec{Vector and Twisted Vector Multiplets}
Consider a compact Lie group $G$ and its Lie algebra ${\frak g}$. Assume that $G$ is a symmetry of the theory which acts non-trivially on chiral multiplets $\Phi$ while leaving the twisted chiral multiplets $\Omega$ invariant. Such a symmetry can be gauged with an ordinary vector multiplet $\CV$, which is a general multiplet of vanishing $R$- and $Z, \t Z$-charges valued in the adjoint representation of ${\frak g}$ and subject to the gauge freedom
\eqn\defCVfd{
\exp{(-2 \CV)} \rightarrow \exp{(i \t\Lambda)} \exp{(-2 \CV)} \exp{(- i \Lambda)}~,
}
with $\Lambda$ and $\t \Lambda$ some arbitrary ${\frak g}$-valued chiral and anti-chiral multiplets of vanishing $R$- and $Z, \t Z$-charges. The expression \defCVfd\ is to be understood in terms of products of general multiplets like in the previous subsection. One can use \defCVfd\ to fix a WZ gauge  
\eqn\compoV{\CV= \big(0,0,0,0,0, a_\mu,\sigma, \t\sigma, \lambda_\pm, \t\lambda_\pm, D\big)~.
}
At first order in the gauge parameters, \defCVfd\ reads 
\eqn\defCVfdii{
\delta_\Lambda \CV = {i\over 2}(\Lambda -\t\Lambda) + {i\over 2} [\Lambda + \t \Lambda, \CV]~,
}
when expanded around \compoV.
The residual gauge transformations are given by $\Lambda= \t\Lambda= (\omega, 0,0, 0)$, which corresponds to
\eqn\resgaugetransfofd{\eqalign{
&\delta_\omega  a_\mu = \d_\mu \omega + i [\omega, a_\mu]~, \qquad
\delta_\omega  \sigma = i [\omega, \sigma]~, \qquad
\delta_\omega  \t\sigma = i [\omega, \t\sigma]~, \cr
&\delta_\omega \lambda_\pm =  i [\omega,\lambda_\pm]~, \qquad\quad
\delta_\omega \t\lambda_\pm =  i [\omega,\t \lambda_\pm]~, \qquad
\delta_\omega D =  i [\omega, D]~.
}}
This identifies $a_\mu$ as a ${\frak g}$-valued gauge field. 
The supersymmetry transformations of $\CV$ in WZ gauge are
\eqn\SusyVector{\eqalign{
\delta a_1 &= - i  \zeta_- \t\lambda_- -i \t\zeta_- \lambda_-~,  \cr
\delta a_{\b 1} &=  i  \zeta_+ \t\lambda_+ +i \t\zeta_+ \lambda_+~,   \cr
\delta \sigma &=  2\zeta_- \t\lambda_+ + 2\t\zeta_+ \lambda_-~, \cr
\delta \t\sigma &=  - 2\t\zeta_- \lambda_+ - 2\zeta_+ \t\lambda_-~, \cr
\delta \lambda_- &= i \zeta_- \left(  D + 2i f_{1\b 1} + \t\CH \sigma  +\half[\sigma, \t\sigma]\right)    + 2 i \zeta_+ D_1 \sigma~, \cr
\delta \lambda_+ &= i \zeta_+ \left(  D - 2i f_{1\b 1} + \CH \t\sigma - \half[\sigma, \t\sigma] \right)  + 2 i \zeta_- D_{\b 1} \t\sigma~, \cr
\delta \t\lambda_- &= -i \t\zeta_- \left(  D - 2i f_{1\b 1}+ \CH \t\sigma   +\half[\sigma, \t\sigma]\right)    - 2 i \t\zeta_+ D_1 \t\sigma~, \cr
\delta \t\lambda_+ &= -i \t\zeta_+ \left(  D + 2i f_{1\b 1} + \t\CH \sigma - \half[\sigma, \t\sigma] \right)   - 2 i\t\zeta_- D_{\b 1} \sigma~, \cr
\delta D &= - 2 D_1\left( \zeta_+ \t\lambda_+ - \t\zeta_+ \lambda_+ \right) + 2 D_{\b 1}\left( \zeta_- \t\lambda_- - \t\zeta_- \lambda_- \right)\cr
&\quad -[\sigma, \zeta_+ \t\lambda_- -\t\zeta_- \lambda_+]
-[\t\sigma, \t\zeta_+ \lambda_- -\zeta_- \t\lambda_+]~.
}}
where we defined the field strength
\eqn\fmunufd{
f_{\mu\nu} =  \d_\mu a_\nu - \d_\nu a_\mu - i [a_\mu, a_\nu]~,
}
and the covariant derivative $D_\mu$ is also gauge-covariant, for instance $D_\mu \lambda_\pm = \nabla_\mu\lambda_\pm - i A_\mu \lambda_\pm - i[a_\mu,\lambda_\pm]$.

We can similarly consider the case when some gauge group $\h G$ only acts non-trivially on twisted chiral multiplets. The corresponding gauge field sits in a twisted vector multiplet $\h \CV$, which is a $\h{\frak g}$-valued general multiplet with vanishing $R$- and $Z,\t Z$-charges subject to the gauge freedom
\eqn\defCVfd{
\exp{(-2 \h\CV)} \rightarrow \exp{(i \t{\h\Lambda})} \exp{(-2 \h\CV)} \exp{(- i \h\Lambda)}~,
}
with $\h\Lambda$ and $\t {\h\Lambda}$ some arbitrary $\h{\frak g}$-valued twisted chiral and twisted antichiral multiplets.
 Let us define the fields
\eqn\deffieldstwistedvec{\eqalign{
&b_1 \equiv  -a_1\, , \quad\qquad\qquad\qquad\; b_{\b 1}\equiv  a_{\b 1}\, ,\qquad\qquad \qquad\quad\; \kappa\equiv  \t M \, , \qquad\quad \t\kappa \equiv  M \, , \cr
& \rho_- \equiv  -\lambda_- - 2i D_1 \t\chi_+\, , \qquad \rho_+ \equiv  \lambda_+ + 2i D_{\b 1}\t\chi_-\, ,\cr
& \t\rho_- \equiv  -\t\lambda_- + 2i D_1 \chi_+\, , \qquad \t\rho_+ \equiv \t \lambda_+ - 2i D_{\b 1}\chi_-~, \qquad  E\equiv  -D -4 D_1 D_{\b 1} C~,
}}
in terms of the components of a $\h{\frak g}$-valued general multiplet of vanishing $R, Z, \t Z$-charges.
Using \defCVfd\ we can fix a WZ gauge where only the components \deffieldstwistedvec\ are nonzero,
\eqn\compoVhat{
\h\CV= \big(b_\mu, \kappa, \t\kappa, \rho_\pm, \t\rho_\pm, E\big)~.}
The fields \deffieldstwistedvec\  transform as
\eqn\resgaugetransfofd{\eqalign{
&\delta_{\h\omega}  b_\mu = \d_\mu \h\omega + i [\h\omega, b_\mu]~, \qquad
\delta_{\h\omega } \kappa = i [\h\omega, \kappa]~, \qquad
\delta_{\h\omega } \t\kappa = i [\h\omega, \t\kappa]~, \cr
&\delta_{\h\omega} \rho_\pm =  i [\h\omega,\rho_\pm]~, \qquad\quad
\delta_{\h\omega} \t\rho_\pm =  i [\h\omega,\t \rho_\pm]~, \qquad
\delta_{\h\omega} E =  i [\h\omega, E]~.
}}
under residual gauge transformation with $\hat\Lambda =  \t {\h\Lambda}=(\h \omega, 0,0,0)$. In particular $b_\mu$ is the gauge field. The supersymmetry transformations in WZ gauge are
\eqn\SusyTwistedVector{\eqalign{
\delta b_1 &= - i  \zeta_- \t\rho_- -i \t\zeta_- \rho_- \, ,   \cr
\delta b_{\b 1} &=  i  \zeta_+ \t\rho_+ +i \t\zeta_+ \rho_+  \, ,   \cr
\delta \kappa &=  -2\zeta_- \rho_+ - 2\zeta_+ \rho_-   \, , \cr
\delta \t\kappa &=   2\t\zeta_- \t\rho_+ + 2\t\zeta_+ \t\rho_- \, , \cr
\delta \rho_- &= i \zeta_- \left(  E + 2i \h f_{1\b 1}  +\half [\kappa, \t\kappa]\right)  -2 i \t\zeta_+ D_1\kappa -i\t\zeta_-\CH \kappa  \, , \cr
\delta \rho_+ &=- i \zeta_+ \left(  E + 2i  \h f_{1\b 1} -\half [\kappa, \t\kappa] \right)  +2 i \t\zeta_- D_{\b 1}\kappa +i\t\zeta_+\t\CH \kappa   \, , \cr
\delta \t\rho_- &=- i \t\zeta_- \left(  E - 2i \h f_{1\b 1}+ \half [\kappa, \t\kappa]\right)    +2 i \zeta_+ D_1\t\kappa + i\zeta_-\t\CH \t\kappa  \, , \cr
\delta \t\rho_+ &= i \t\zeta_+ \left(  E - 2i \h f_{1\b 1} -\half [\kappa, \t\kappa]\right)  -2 i \zeta_- D_{\b 1} \t\kappa -i \zeta_+\CH \t \kappa  \, , \cr
\delta E &=  2 D_1\left( \zeta_+ \t\rho_+ - \t\zeta_+ \rho_+ \right) + 2 D_{\b 1}\left( \zeta_- \t\rho_- - \t\zeta_- \rho_- \right) \cr
&\qquad +[\kappa, \t\zeta_+ \t\rho_-  -\t\zeta_-\t\rho_+] + [\t\kappa, \zeta_+\rho_- - \zeta_-\rho_+]~,
}}
where we defined the field strength 
\eqn\fmunufd{
\h f_{\mu\nu} =  \d_\mu b_\nu - \d_\nu b_\mu - i [b_\mu, b_\nu]~,
}
and $D_\mu$ is also gauge-covariant.
Note that the scalars $\kappa$, $\t\kappa$ in the twisted vector multiplet have $R$-charge $\pm 2$, respectively.

\subsec{Field Strength Multiplets}
In the case of an Abelian vector multiplet or twisted vector multiplet, one can define interesting gauge-invariant field strength multiplets. Given an Abelian vector multiplet \compoV, we can define the twisted chiral multiplet 
\eqn\SigmaTwistedChiral{
\Sigma= \left(\omega, \,\eta_-,\, \t\eta_+,\, G\right) = \left(\sigma , \, \sqrt2 \lambda_- , \, -\sqrt2 \t\lambda_+ , \, i D - 2 f_{1\b 1} + i \t\CH \sigma\right)~,
}
which in flat space superfield notation reads $\Sigma = - i D_- \t D_+ \CV$, 
and the twisted antichiral multiplet
\eqn\SigmaTwistedAntiChiral{
\t\Sigma= \left(\t\omega, \,\t\eta_-,\, \eta_+,\, \t G\right) = \left(\t \sigma , \, \sqrt2 \t\lambda_- , \, -\sqrt2 \lambda_+ , \, -i D - 2 f_{1\b 1} - i \CH \t\sigma\right)~,
}
which in flat space reads $\t\Sigma = i D_+ \t D_- \CV$.
 
Similarly, given an Abelian twisted vector multiplet \compoVhat\ we can define the chiral multiplet 
\eqn\HatSigmaChiral{
\h\Sigma= \left(\phi, \,\psi_-,\, \psi_+,\, F\right) = \left(\kappa , \, -\sqrt2 \rho_- , \, \sqrt2  \rho_+ , \,- i E + 2 \h f_{1\b 1}\right)~,
}
which in flat space reads $\h\Sigma = -i \t D_- \t D_+ \h\CV$, and the antichiral multiplet
\eqn\HatSigmaAntiChiral{
\t{\h\Sigma}= \left(\t\phi, \,\t\psi_-,\, \t\psi_+,\, \t F\right) = \left(\t\kappa , \, -\sqrt2 \t\rho_- , \, \sqrt2  \t\rho_+ , \, i E + 2 \h f_{1\b 1}\right)~,
}
which in flat space reads $\t{\h\Sigma} = i D_+ D_- \h\CV$.
Note that $\h\Sigma$ and $\t {\h \Sigma}$ have $R$-charge $\pm 2$ and vanishing central charge.

\subsec{Charged Chiral and Twisted Chiral Multiplets}
Consider a chiral  multiplet $\Phi$ and an antichiral multiplet $\t\Phi$ in some representation ${\frak R}$ and $\b{\frak R}$  of the gauge algebra ${\frak g}$, respectively, with the infinitesimal gauge transformations
\eqn\actiononchiral{
\delta_\Lambda \Phi =i\Lambda \Phi~,\qquad
\delta_\Lambda \t\Phi = -i \t\Phi \t\Lambda~.
}
Here $\Lambda$ is a ${\frak R}$-valued chiral multiplet and $\t\Lambda$ a $\b{\frak R}$-valued antichiral multiplet of vanishing $R$- and $Z, \t Z$-charges. The supersymmetry transformations of $\Phi$ minimally coupled to a vector multiplet $\CV$ in WZ gauge are
\eqn\ChiralSusyWZ{\eqalign{
\delta \phi &= \sqrt{2} (\zeta_+\psi_- -\zeta_-\psi_+)~,\cr
\delta \psi_- &= \sqrt{2} \zeta_- F - i\sqrt{2}  \t\zeta_- \left( z-\sigma -{r\over 2} \CH\right)  \phi +2i \sqrt2 \t\zeta_+ D_1\phi~,\cr
\delta \psi_+ &= \sqrt{2} \zeta_+ F - i\sqrt{2}   \t\zeta_+  \left(\t z-\t\sigma-{r\over 2} \t\CH\right)\phi +2i \sqrt2 \t\zeta_- D_{\b 1}\phi~, \cr
\delta F &= \sqrt2 i  \left(\t z-\t\sigma-{r\over 2} \t\CH\right) \t\zeta_+\psi_- - \sqrt2 i  \left( z-\sigma-{r\over 2} \CH\right)  \t\zeta_-\psi_+~, \cr
& \quad +2i \sqrt2 \t\zeta_+ D_1\psi_+ - 2i \sqrt2 \t\zeta_-D_{\b 1}\psi_- + 2 i (\t\zeta_+ \t\lambda_- -\t\zeta_- \t\lambda_+)\phi~,
}}
where $D_\mu$ is gauge covariant and the vector multiplet fields $(a_\mu, \sigma, \t\sigma, \lambda_\pm, \t\lambda_\pm)$ are ${\frak R}$-valued. Similarly, for $\t\Phi$ we have
\eqn\AntiChiralSusyWZ{\eqalign{
\delta \t\phi &= -\sqrt{2} (\t\zeta_+\t\psi_- -\t\zeta_-\t\psi_+)~,\cr
\delta \t\psi_- &= \sqrt{2} \t\zeta_- \t F + i\sqrt{2} \zeta_-   \t\phi \left(\t z-\t\sigma-{r\over 2} \t\CH\right) -2i \sqrt2 \zeta_+ D_1\t\phi~,\cr
\delta \t\psi_+ &= \sqrt{2} \t\zeta_+ \t F + i\sqrt{2} \zeta_+ \t\phi  \left( z-\sigma -{r\over 2} \CH\right) -2i \sqrt2 \zeta_- D_{\b 1}\t\phi~,\cr
\delta \t F &= \sqrt2 i \, \zeta_+\t\psi_-\left(z-\sigma-{r\over 2}\CH\right) - \sqrt2 i \,  \zeta_-\t\psi_+ \left( \t z-\t\sigma-{r\over 2} \t\CH\right)\cr
& \quad +2i \sqrt2 \zeta_+D_1 \t\psi_+ - 2i \sqrt2 \zeta_-D_{\b 1}\t\psi_-
+ 2 i\t\phi (\zeta_+ \lambda_- -\zeta_- \lambda_+)~,
}}
with $\b{\frak R}$-valued vector multiplet fields.

Consider also a twisted chiral  multiplet $\Omega$ and a twisted antichiral multiplet $\t\Omega$ in the representations ${\frak R}$ and $\b{\frak R}$  of the gauge algebra $\h{\frak g}$, with the infinitesimal gauge transformations
\eqn\actionontwistedchiral{
\delta_{\h\Lambda} \Omega =i\h\Lambda \Omega ~,\qquad
\delta_{\h\Lambda} \t\Omega = -i \t\Omega {\t{\h\Lambda}}~.
}
Here $\h\Lambda$  is a ${\frak R}$-valued twisted chiral multiplet and $\t{\h\Lambda}$ a $\b{\frak R}$-valued twisted antichiral multiplet of vanishing $R$- and $Z, \t Z$-charges. The supersymmetry transformations of $\Omega$ minimally coupled to a twisted vector multiplet $\h\CV$ in WZ gauge are
\eqn\TwistedChiralGaugedSusy{\eqalign{
\delta \omega &= \sqrt{2} (\t\zeta_+\eta_- - \zeta_-\t\eta_+)~, \cr
\delta \eta_- &= \sqrt{2} \zeta_- G  -i\sqrt2 \t\zeta_- \kappa \omega  +2i\sqrt{2}  \zeta_+ D_1\omega~, \cr
\delta \t\eta_+ &= \sqrt{2} \t\zeta_+ G   -i\sqrt2  \zeta_+ \t\kappa \omega +2i \sqrt{2}  \t\zeta_-D_{\b 1}\omega~, \cr
\delta G &= 2i \sqrt{2}  (\zeta_+ D_1 \t\eta_+ - \t\zeta_- D_{\b 1} \eta_-)+ i\sqrt2 (\zeta_+ \t\kappa \eta_- - \t\zeta_- \kappa \t\eta_+) + 2i  ( \zeta_+ \t\rho_- - \t\zeta_- \rho_+)\omega~,
}}
where $D_\mu$ is also gauge covariant (including the gauge field $b_\mu$) and the twisted vector multiplet fields $(b_\mu, \kappa, \t\kappa, \rho_\pm, \t\rho_\pm)$ are ${\frak R}$-valued. Similarly, for the twisted antichiral multiplet:
\eqn\TwistedAntiChiralGaugedSusy{\eqalign{
\delta \t\omega &= -\sqrt{2} (\zeta_+\t\eta_- - \t\zeta_-\eta_+)~, \cr
\delta \t\eta_- &= \sqrt{2} \t\zeta_- \t G   +i\sqrt2 \zeta_- \t\omega \t\kappa  -2i\sqrt{2}  \t\zeta_+ D_1\t\omega~, \cr
\delta \eta_+ &= \sqrt{2} \zeta_+ \t G    +i\sqrt2  \t\zeta_+ \t\omega  \kappa  -2i \sqrt{2}  \zeta_-D_{\b 1}\t\omega~, \cr
\delta \t G &= 2i \sqrt{2}  (\t\zeta_+ D_1 \eta_+ - \zeta_- D_{\b 1} \t\eta_-)
+ i\sqrt2 ( \t\zeta_+  \t\eta_- \kappa -\zeta_- \eta_+\t\kappa ) + 2i  \,\t\omega ( \t\zeta_+ \rho_- - \zeta_- \t\rho_+)~,
}}
with $\b{\frak R}$-valued twisted vector multiplet fields.

\subsec{Supersymmetric Lagrangians}
Using the above results, it is straightforward to construct  supersymmetric actions of the form
\eqn\susyaction{S =\int_{\Sigma} d^2x \sqrt{g}\,{\scr L}~,}
which generalize the usual flat space formulas to the case of rigid supersymmetry on a Riemann surface $\Sigma$. We have the following possibilities:

\medskip

\item{1.)} {\it D-Terms.} It is clear from \SusyCplxSuperfii\ that the $D$-term of any general multiplet $\CS$ of vanishing $R$- and $Z, \t Z$-charge is a good supersymmetric Lagrangian,
\eqn\LagD{
{\scr L}_D = D~.
}

\item{2.)} {\it F-Terms.}
 Given any chiral multiplet $\Phi$ with $r=2$ and $z=\t z=0$, and an anti-chiral multiplet $\t\Phi$ with $r=-2$ and $z=\t z=0$, we have
\eqn\LagF{
{\scr L}_F = F +\t F\, .
}

\item{3.)} {\it Twisted F-Terms ($G$-Terms).}
Similarly, given any twisted chiral multiplet $\Omega$ and twisted antichiral multiplet $\t\Omega$, the Lagrangian
\eqn\LagG{
{\scr L}_G = \left(G - i \t \CH \omega \right) + \left(\t G + i \CH \t\omega \right) 
}
is supersymmetric. Such $G$-terms are special to two dimensions and play an important role in curved space backgrounds.

\item{4.)} {\it Improvement of the $\CR$-multiplet.} Another interesting supersymmetric Lagrangian can be obtained by coupling the $R$-symmetry gauge field $A_\mu$ to a conserved current sitting in a linear multiplet $\CJ$.  Using \LinearMultSusy\ and the integrability condition for the Killing spinor equation (see Appendix B), one can see that the following Lagrangian is supersymmetric: 
\eqn\LagImprovR{
{\scr L}_{\CJ} = A_\mu j^\mu + {1\over 4} \t \CH K + {1\over 4} \CH \t K  - {1\over 4} R J\, , 
}
where $R$ is the Ricci scalar. This Lagrangian is special to curved space supersymmetry (it vanishes in flat space). Note that the conserved current $j^\mu$ in \LagImprovR\ should be conserved off-shell. However, the Lagrangian \LagImprovR\ also appears as an improvement of the $R$-symmetry current by an ordinary (on-shell) conserved current, $j_\mu^{(R)}\rightarrow j_\mu^{(R)}+ \Delta r\, j_\mu$, and it is supersymmetric at first order in the deformation parameter  $\Delta r$ (seagull terms are needed at second order by gauge invariance).  Note also that there is no similar supersymmetric completion of $A_\mu \h j^\mu$, with $\h j^\mu$ the conserved current of a twisted linear multiplet $\h \CJ$, consistent with the fact that we cannot improve the $R$-symmetry current by an axial symmetry (i.e. a symmetry acting on twisted chiral multiplets) without also violating the $\CR$-multiplet constraints \Rmultdefi\ \Efratwip.%
\foot{This is the same as saying that we need $r=0$ for our curved-space twisted chiral multiplets, a fact we  mentioned in section 6.3.}

\medskip

From the above rules we can directly work out all the standard Lagrangians. The canonical kinetic term for a chiral multiplet $\Phi$ of charges $r, z, \t z$, coupled to a vector multiplet $\CV$ like in section 6.8, can be extracted from the $D$-term of $-\half \Tr\, \t\Phi e^{-2\CV} \Phi$, with lowest component $-\half \Tr\, \t\phi\phi$ in Wess-Zumino gauge.
One finds:
\eqn\LagPhi{\eqalign{
{\scr L}_{\t\Phi\Phi} = &\;   2 D_1\t\phi D_{\b 1} \phi+ 2 D_{\b 1}\t\phi D_{1} \phi-  \t F F  +2  i \t\psi_+ D_1\psi_+ -2 i \t\psi_- D_{\b 1}\psi_-  + \t\phi D\phi  \cr 
&-\left({r\over 4} R -\half \CH \t z -\half \t\CH z\right) \t\phi\phi   +\t\phi\left(\t z-\t\sigma -{r\over 2}\t\CH\right) \left(z-\sigma -{r\over 2}\CH\right)\phi  \cr
&+\half \t\phi [\sigma, \t\sigma]\phi+ i  \t\psi_+\left(\t z-\t\sigma -{r\over 2}\t\CH\right)\psi_- -i  \t\psi_-\left(z-\sigma -{r\over 2}\CH\right) \psi_+\cr
& +i \sqrt2 \big(\t\psi_+\t\lambda_- -\t\psi_-\t\lambda_+\big)\phi + i\sqrt2 \, \t\phi \big(\lambda_+\psi_- -\lambda_-\psi_+\big)~.
}}
Here $R$ is the Ricci scalar, the covariant derivatives are also gauge covariant (with the gauge field $a_\mu$) and the overall trace over the gauge group is implicit.
We can also have superpotential terms like in flat space. For any quasi-homogeneous holomorphic function $W$ of the chiral multiplets, of $R$-charge $2$ and vanishing $Z, \t Z$-charges (and similarly for the anti-chiral multiplets), we have
\eqn\LagW{{\scr L}_{W+ \t W} = F^i \d_i W + \psi^i_- \psi^j_+ \d_i \d_j W  
				+\t F^i \d_i \t W - \t \psi^i_- \t\psi^j_+ \d_i \d_j \t W}
Combining \LagPhi\ with \LagW, we see that the superpotential contributes $\d_i \t W \d^i W$ to the scalar potential, like in flat space.

The kinetic Lagrangian for the twisted chiral multiplet coupled to a twisted vector multiplet $\h\CV$ is given by the $D$-term of $\half\Tr\,  \t\Omega  e^{-2\h\CV}\Omega$, with lowest component $\half \Tr\, \t\omega\omega$ in Wess-Zumino gauge. It reads
\eqn\LagOmegaTwisted{\eqalign{
{\scr L}_{\t\Omega\Omega} =&\;  2 D_{\b 1} \t\omega D_1\omega+2 D_{1} \t\omega D_{\b 1}\omega - \t G G +2 i\, \eta_+ D_1 \t\eta_+    -2i\, \t\eta_- D_{\b 1} \eta_- 
  \cr
& +\half \t\omega \big(\kappa\t\kappa+\t\kappa \kappa\big) \omega +\t\omega E \omega - i \t\eta_-\kappa \t\eta_+ + i \eta_+ \t\kappa \eta_-  \cr
&+ i\sqrt2 \big(\eta_+\t\rho_- -\t\eta_- \rho_+\big)\omega + i\sqrt2 \,\t\omega \big(\t\rho_+\eta_- -\rho_- \t\eta_+\big)~, }}
where the covariant derivatives are also gauge-covariant (with the gauge field $b_\mu$) and the trace over the gauge group is implicit.
Using \TwistedchiralmultW\ and \LagG\ we can also construct twisted superpotential couplings from any holomorphic function $\hat W$ and any antiholomorphic function $\t {\hat W}$:
\eqn\LagWtwist{
\CL_{\hat W+ \t{\hat W}} =  G^i \d_i {\hat W} +\eta_-^i \t\eta_+^j \d_i \d_j {\hat W} + i \t \CH \hat W 
				+\t G^i \d_i \t{\hat W} -\t\eta_-^i \eta_+^j \d_i \d_j \t {\hat W} - i \CH \t{\hat W}\, .
}
Note that the twisted superpotential  Lagrangian contains an extra contribution $i \t\CH W - i \CH \t{\hat W}$ with respect to its flat space expression.

The Yang-Mills Lagrangian for the ordinary vector multiplet \compoV\ in WZ gauge is obtained by extracting the $D$-term of the gauge-invariant general multiplet of lowest component ${1\over 4}\Tr\left( \t\sigma\sigma\right)$. This gives
\eqn\LagVectorYM{\eqalign{
{\scr L}_{\CV} = &\, \half \left(2i f_{1\b 1}+\half \t\CH \sigma -\half \CH \t\sigma \right)^2 + D_{\b 1}\t\sigma D_1\sigma+ D_1 \t\sigma D_{\b 1}\sigma +{1\over 8} [\sigma, \t\sigma]^2 \cr
&\,        + 2i \t\lambda_+ D_1\lambda_+ -2 i \t\lambda_- D_{\b 1}\lambda_-
+ i \t\lambda_- [\sigma, \lambda_+] - i \t\lambda_+ [\t\sigma, \lambda_-] \cr
&\,
 -\half \left( D+\half \t\CH \sigma +\half \CH \t\sigma  \right)^2~,
}}
where $f_{\mu\nu}$ is the field strength defined in \fmunufd. The trace over the gauge group and the overall gauge coupling ${1\over g^2}$ are implicit. For a $U(1)$ vector multiplet, we can also consider the Fayet-Iliopoulous (FI) coupling, which is the $G$-term of the field strength multiplet \SigmaTwistedChiral, \SigmaTwistedAntiChiral:
\eqn\FIterm{
{\scr L}_{\Sigma}=  \xi D  + i {\theta \over 2\pi } \, 2i f_{1\b 1}
}
Here $\xi$ is the FI parameter and $\theta$ is the topological angle (of period $2\pi$). They pair into a holomorphic coupling $\tau = {\theta\over 2\pi} - i\xi $, which can be viewed  as the bottom component of a background twisted chiral multiplet. (The FI term \FIterm\ is a special case of the twisted superpotential \LagWtwist\ with $\h W= \half \tau \Sigma$.)

The Yang-Mills Lagrangian for a  twisted vector multiplet \compoVhat\ in WZ gauge is obtained by extracting the $D$-term of the gauge invariant general multiplet of lowest component $-{1\over 4}\Tr(\t\kappa\kappa)$. We obtain 
\eqn\LagTwistedVector{\eqalign{
{\scr L}_{\hat \CV} = &\; \half \left(2i \h f_{1\b 1}\right)^2 + D_{\b 1}\t\kappa D_1\kappa +D_{1}\t\kappa D_{\b 1}\kappa  
 + 2i \t\rho_+ D_1\rho_+ -2 i \t\rho_- D_{\b 1}\rho_-     \cr
          &\, -\half E^2 +{1\over 8}[\kappa,\t\kappa]^2 -{1\over 4}\left( R-2\CH\t\CH\right)\t\kappa\kappa  -i \CH \, \t\rho_-\rho_+  + i \t \CH \,\t\rho_+\rho_- \cr
&\; -i\t\rho_+[\kappa,\t\rho_-]+ i \rho_+[\t\kappa, \rho_-]~,
}}
where $\h f_{1\b 1}$ is the field strength defined in \fmunufd, $R$ is the Ricci scalar, and the trace over the gauge group and the overall gauge coupling ${1\over{\h g}^2}$ are implicit.
Note that the coupling to curved space generally induces a mass term for the complex scalar in the twisted vector multiplet. However, in the case of a background preserving four supercharges this effective mass vanishes. 
For a $U(1)$ twisted vector multiplet, we also have the twisted FI parameter, which is the $F$-term of the twisted field strength chiral multiplet \HatSigmaChiral, \HatSigmaAntiChiral:
\eqn\FItermtwist{
{\scr L}_{\h\Sigma}=  \h\xi E  + i {\h \theta \over 2\pi } \, 2i \h f_{1\b 1}~.
}
The twisted FI parameter and $\h\theta$ angle are paired as $\h\tau={\h\theta\over 2\pi}- i \h\xi $, which can be viewed  as the bottom component of a background chiral multiplet. (The FI term \FItermtwist\ is a special case of the superpotential \LagW\ with $W= \half \h\tau \h\Sigma$.)

Finally, consider the improvement Lagrangian \LagImprovR\ with a linear multiplet $\CJ=  \CF+ \t{\CF}$, with $ \CF= \CF(\Omega^i)$ a holomorphic function of twisted chiral multiplets and  $ \t{ \CF}= \t{ \CF}(\t\Omega^i)$ an antiholomorphic function of twisted antichiral multiplets. We find the supersymmetric Lagrangian
\eqn\improvhW{\eqalign{
{\scr L}_{ \CF + \t{ \CF}} = &\; 2 i A_1\, \d_{\b 1}( \CF-\t { \CF})- 2 i A_{\b 1}\, \d_{1}( \CF-\t { \CF}) 
 \,- {1\over 4} R \big( \CF+ \t{ \CF}\big)  \cr
&+ {i\over 2} \t \CH \left(\t G^i \d_i \t{ \CF} + \eta^i_+ \t\eta^j_- \d_i\d_j  \t{ \CF}\right)- {i\over 2} \CH  \left(G^i \d_i  \CF + \eta^i_- \t\eta^j_+ \d_i\d_j   \CF\right)~, 
}}
with $R$ the Ricci scalar.
Note that \improvhW\ leads to a dimensionless action whenever $\omega^i, \t\omega^i$ are themselves dimensionless. This provides an interesting finite counterterm on any Riemann surface of nonzero curvature ---in particular on the sphere \GerchkovitzGTA.%
\foot{We thank Zohar Komargodski for interesting discussions on this point.}

\subsec{Supersymmetric Non-Linear Sigma Models}

Finally, using the formula \DofK\ we can write down the curved-space supersymmetric Lagrangian for any non-linear sigma model with flat-space Lagrangian
\eqn\flatL{
{\scr L}_{K(\t\Phi,\Phi)} = \int d^4\theta\, K(\t \Phi,\Phi)~, \qquad\qquad
{\scr L}_{K(\t\Omega,\Omega)} = \int d^4\theta\, K(\t\Omega, \Omega)~,
}
for some chiral multiplets $\Phi^i$ 
or twisted chiral multiplets $\Omega^n$. The curved-space Lagrangians presented below generalize the ones given in \JiaFOA\ for the round $S^2$ without R-symmetry flux.~\foot{We could also consider more general sigma models involving both chiral and twisted chiral multiplets, as well as the semi-chiral multiplets of Appendix E ---see for instance \LindstromZR\ for a discussion in flat space. We restrict ourselves to \flatL\ for simplicity. }

Consider first a theory of chiral multiplets $\Phi^i$ with charges $r^i, z^i, \t z^i$ and antichiral multiplets $\t \Phi^{\b i}$ of charges $r^{\b i}= -r^i, z^{\b i}= -z^i, \t z^{\b i}=-\t z^i$. 
Defining the K\"ahler metric on target space
\eqn\Kmet{
g_{i\b j}= K_{i\b j}~,
}
the corresponding non-vanishing Christoffel symbols are given by
\eqn\Christsymb{
\Gamma^k_{ij} = g^{k\b l} K_{ij \b l}~,\qquad\quad
\Gamma^{\b k}_{\b i\b j} = g^{l \b k} K_{\b i\b j  l}~.
}
We also denote ordinary derivatives of the metric by a comma, for instance $g_{i\b j, k}= K_{i\b j k}$.
Up to a total derivative, the Lagrangian is given by $-\half D^{\CK}$:
\eqn\dchiral{\eqalign{
{\scr L}_{K(\t\Phi,\Phi)} =& \; 
g_{i\jbar}\,\Big(2 D_1 \phi^i D_{\b 1}\t\phi^{\jbar}
+2 D_{\b 1} \phi^i D_{ 1}\t\phi^{\jbar} - F^i \t F^{\jbar} -\half \big(z^i_\CH \t z^{\jbar}_\CH+ z^{\jbar}_\CH \t z^{i}_\CH\big)\phi^i \t\phi^{\jbar} \Big)\cr
&\; -\half K_i \,\Big( {r^i \over 4} R -\half \big(z^i \t\CH +\t z^i \CH\big) \Big)\phi^i 
+\half K_{\b i} \,\Big( {r^{\b i} \over 4} R -\half \big(z^{\b i} \t\CH +\t z^{\b i} \CH\big)  \Big)\t \phi^{\b i}\cr
&\; + g_{i\b j} \left(2 i \t\psi_+^{\b j} {\bf D}_1 \psi_+^i - 2 i \t\psi_-^{\b j} {\bf D}_{\b 1} \psi_-^i \right)\cr
&\; -{i\over 2} \t\psi_-^{\b j}\psi_+^i \left( g_{l\b j} \nabla_i(z_\CH \phi)^l - g_{i\b l}\nabla_{\b j}(z_\CH \t\phi)^{\b l}\right)\cr
&\; +{i\over 2} \t\psi_+^{\b j}\psi_-^i \left( g_{l\b j} \nabla_i(\t z_\CH \phi)^l - g_{i\b l}\nabla_{\b j}(\t z_\CH \t\phi)^{\b l}\right) \cr
&\; - g_{i\b k, j}\,\psi_-^i \psi_+^j \t F^{\b k} +  g_{k\b i, \b j}\,\t\psi^{\b i}_- \t\psi_+^{\b j}  F^{ k}
+ g_{i\b j, k\b l}\, \psi^i_+ \psi^k_- \t\psi^{\b k}_+ \t\psi^{\b l}_-~,
}}
where we have defined the covariant derivatives
\eqn\defops{\eqalign{
&  {\bf D}_{\mu} \psi_\pm^i =  D_\mu \psi^i_\pm + \Gamma^i_{jk} (D_\mu \phi^k) \psi^j_\pm~,\cr
& \nabla_i X^j = \d_i X^j + \Gamma^j_{ik}X^k~,\cr
& \nabla_{\b i} \t X^{\b j} = \d_{\b i}\t X^{\b j} + \Gamma^{\b j}_{\b i\b k}\t X^{\b k}
}}
and used the notation $(z_\CH \phi)^l \equiv z_\CH^l \phi^l$, etc.

Note the presence in \dchiral\ of the extra terms in $ K_i, K_{\b i}$ arising from coupling to curved space (or from the central charge, which could be there in flat space). The quasi-homogeneity conditions \quasihom\ ensure that the curved space Lagrangian \dchiral\ is invariant under K\"ahler transformations
\eqn\Ktransfo{
K(\t\phi^{\ibar}, \phi^i) \rightarrow K(\t\phi^{\ibar}, \phi^i) + f(\phi^i)+ \t f(\t\phi^{\ibar})~,
}
as in flat space.%
\foot{We thank Guido Festuccia for useful discussions on this point.}

Similarly, the Lagrangian of a non-linear sigma model of twisted chiral multiplets $\Omega^n$ and twisted antichiral multiplets $\t\Omega^{\b n}$, which have vanishing charges, is given by $\half D^{\CK}$. We introduce the K\"ahler metric on target space
\eqn\Kmettc{
g_{m\b n}= K_{m\b m}~,
}
with non-vanishing Christoffel symbols
\eqn\Christsymbtc{
\Gamma^p_{mn} = g^{p\b q} K_{mn \b q}~,\qquad\quad
\Gamma^{\b p}_{\b m\b n} = g^{\b p q} K_{\b m\b n  q}~.
}
Up to a total derivative, we obtain the non-linear sigma model Lagrangian
\eqn\dtwistedchiral{\eqalign{
{\scr L}_{K(\t\Omega, \Omega)} =&\; 
g_{m\b n}\, \Big( 2 D_1 \omega^m D_{\b 1}\t\omega^{\b n}+2 D_{\b 1} \omega^m D_{ 1}\t\omega^{\b n} - G^m \t G^{\b n}\Big)\cr
&\; + g_{m\b n} \left(2 i \eta_+^{\b n} {\bf D}_1 \t\eta_+^m - 2 i \t\eta_-^{\b n} {\bf D}_{\b 1} \eta_-^m \right)\cr
&\; - g_{m\b n, p}\,\eta_-^m \t\eta_+^p \t G^{\b n} +  g_{m\b n, \b p}\,\t\eta_-^{\b n} \eta_+^{\b p}  G^m - g_{m\b n, \b p q}\, \t\eta_+^m \t\eta_-^{\b n} \eta_+^{\b p} \eta_-^q  ~,
}}
where 
\eqn\defopstc{\eqalign{
&  {\bf D}_{\mu} \t\eta_+^m =  D_\mu \t\eta_+^m + \Gamma^m_{np} (D_\mu \omega^p) \t\eta_+^n~,\cr
& {\bf D}_{\mu} \eta_-^m =  D_\mu \eta_-^m + \Gamma^m_{np} (D_\mu \omega^p) \eta_-^n~.
}}
The non-linear sigma model Lagrangian for twisted chiral multiplets \dtwistedchiral\ is invariant under K\"ahler transformations
\eqn\Ktransfotc{
K( \t\omega^{\b n}, \omega^n) \rightarrow K( \t\omega^{\b n}, \omega^n) + g(\omega^n)+ \t g(\t\omega^{\b n})~.
}

\vskip 1cm

\noindent {\bf Acknowledgments:}
We would like to thank Francesco Benini, Thomas Dumitrescu, Guido Festuccia,  Efrat Gerchkovitz, Zohar Komargodski, Daniel Park, Martin Ro\v cek and Itamar Shamir for stimulating discussions and comments.
A preliminary version of these results was presented by SC at SUSY 2013 at ICTP Trieste (26-31 August 2013). 
The work of SC is supported in part by the STFC Consolidated Grant ST/J000353/1.
Any opinions, findings, and conclusions or recommendations expressed in this
material are those of the authors and do not necessarily reflect the views of the funding agencies.

\appendix{A}{Conventions}
\subsec{Flat Space Conventions}
We work in Euclidean signature.%
\foot{Our flat-space conventions are the analytical continuations to the Euclidean of the ones of \DumitrescuIU\ (Appendix C). In particular vectors are analytically continued as  $(X^1,X^2)=(X^1, i X^0)$, and covectors as $(X_1,X_2)=(X_1, -i X_0)$.}
The flat-space metric is $\delta_{\mu\nu}$, $\mu,\nu=1,2$ and the Levi-Civita symbol $\epsilon^{\mu\nu}$ is normalized to $\epsilon^{12}=1$. We mostly work in complex coordinates $z= x^1 + i x^2, \bz = x^2 - i x^2$, in which case $\delta_{z\bz}= \half$, $\delta_{zz}=\delta_{\bz\bz}=0$ and $\epsilon^{z\bz}= -2 i$.  
In particular, any covector $X_\mu$ is decomposed to
\eqn\covec{
X_z = \half (X_1- i X_2)~, \qquad  X_{\b z} = \half (X_1 + i X_2)~.
}
One may call the holomorphic and antiholomorphic components $X_z$ and $X_\bz$ the left-moving and right-moving components, respectively.

The minimal spinors in two Euclidean  dimensions are the Weyl spinors $\psi_-$ and $\psi_+$ of spin $\half$ and $-\half$, respectively, under ${\rm Spin}(2) \cong U(1)$. It is sometimes useful to consider Dirac spinors
\eqn\Diracspin{
\psi = (\psi_\alpha)= \left(\matrix{\psi_- \cr \psi_+}\right)~.
}
Our conventions for the two-dimensional gamma matrices are ${(\gamma^{\mu})^{\alpha}}_\beta={ (-\sigma^1, -\sigma^2)^{\alpha}}_\beta$ when $\mu$ runs over $x^1, x^2$, and $\gamma^3 = \sigma^3$, with $\sigma^a$ the Pauli matrices. They satisfy $\gamma^\mu \gamma^\nu = \delta^{\mu\nu}+i \epsilon^{\mu\nu}\gamma^3$ and $\{ \gamma^3,\gamma^\mu\}=0$.
In complex coordinates, we have
\eqn\gammazzb{
\gamma_z= \left(\matrix{0 & 0 \cr -1 & 0}\right)~, \qquad \gamma_\bz= \left(\matrix{0 & -1 \cr 0 & 0}\right)~.}
Dirac indices are raised and lowered with the epsilon symbols $\epsilon^{\alpha\beta}, \epsilon_{\alpha\beta}$ and are contracted from upper-left to lower-right in the usual way, so that
\eqn\expprod{
\psi \chi = \psi_+ \chi_- -\psi_- \chi_-~, \qquad
\psi \gamma^3\chi = \psi_+ \chi_- +\psi_- \chi_-~.
}
We could also write the covector $\covec$ as a bispinor
\eqn\bispinor{
  X_{--}= -4 X_z~, \qquad  X_{++}= 4 X_{\b z}~,
}
which manifests the fact that $X_z$ and $X_\bz$ are objects of definite spin $\pm 1$ in flat space.

Flat space superfields are functions of the superspace coordinates $(z,\bz, \theta^\pm, \t\theta^\pm)$. The vector $R$-charge of $\theta^\pm$ and $\t\theta^\pm$ are $\pm 1$, respectively. 
The supercharges $Q_\pm, \t Q_\pm$ act on superspace as
\eqn\defDpii{\eqalign{
& Q_+ = {\d \over \d \theta^+} + 2 i \t \theta^+ \del_{\b z}~, \qquad \t Q_+ = - {\d \over \d \t \theta^+} - 2 i  \theta^+ \del_{\b z}~,  \cr
&Q_- = {\d \over \d \theta^-} - 2 i \t \theta^- \del_z~, \qquad \t Q_- = - {\d \over \d \t \theta^-} + 2 i  \theta^- \del_z~.
}}
The supersymmetry covariant derivatives read
\eqn\defDpii{\eqalign{
& D_+ = {\d \over \d \theta^+} - 2 i \t \theta^+ \del_{\b z}~, \qquad \t D_+ = - {\d \over \d \t \theta^+} + 2 i  \theta^+ \del_{\b z}~,  \cr
&D_- = {\d \over \d \theta^-} + 2 i \t \theta^- \del_z~, \qquad \t D_- = - {\d \over \d \t \theta^-} - 2 i  \theta^- \del_z~,
}}
whose non-vanishing anticommutators are 
$\{D_- , \t D_- \} = - 4 i \d_z$ and $\{D_+ , \t D_+ \} =  4 i \d_{\b z}$.

\subsec{Curved Space Conventions}
Consider an orientable two-manifold $\Sigma$ with a Riemannian metric $g_{\mu\nu}$. 
Any such manifold is Hermitian and K\"ahler. The complex structure ${J^\mu}_\nu$ is given explicitly by the Levi-Civita tensor, $J_{\mu\nu}= -\epsilon_{\mu\nu}$, which is a closed two-form.
We therefore consider $\Sigma$ with  K\"ahler metric
\eqn\genmetrici{
ds^2= 2 g_{z\bz}(z, \bz) dz d\bz~.}
To describe spinors, we introduce a complex frame
\eqn\choiceframe{
e^1 = g^{1\over 4} dz~, \qquad  e^{\b 1} = g^{1\over 4} d\bz~, 
}
where $g$ is the determinant of the metric defined  through $\sqrt{g}= 2 g_{z\bz}(z, \bz)$ (the factor of $2$ in this definition simplifies some formulas).
This is the most natural choice of frame in two dimensions and we always use it throughout this paper. For simplicity of notation, we often write all tensors including covariant derivatives in the frame basis. Frame indices are raised and lowered with $\delta^{1\b 1}= 2$, $\delta_{1\b 1}=\half$.

The non-zero Christoffel symbols for the Levi-Civita connection of the metric \genmetrici\ are
\eqn\LCconn{
\Gamma_{z z}^z =\half \d_z \log g~, \qquad \Gamma_{\bz \bz}^\bz =\half \d_\bz \log g~.
}
The corresponding spin connection $\omega_\mu$ reads
\eqn\spincon{
 \quad \omega_z = -{i\over 4} \d_z \log g~, \qquad 
 \quad \omega_{\bz }= {i\over 4} \d_\bz \log g~,
}
which is really an Abelian connection on the $U(1)$ spin bundle. We defined $\omega_\mu \equiv - 2 i \omega_{\mu 1\b 1}$, where $\omega_{\mu ab}$ is the spin connection in any frame $\{e^a\}$. 
The Riemann tensor $R_{\mu\nu\rho\sigma}$ has only one independent component. In complex coordinates,
\eqn\Riemanntensor{
R_{z\bz z\bz} = i g_{z\bz} \,(\d_z \omega_\bz-\d_\bz \omega_z)= -\half g_{z\bz} g_{z\bz} R~, \qquad\quad  R= {2\over \sqrt{g}} \d_z\d_\bz \log g~,}
with  $R$ the  Ricci scalar. Note that in our conventions $R= -2$ on the round sphere of unit radius.
The covariant derivative on any field $\varphi_{(s)}$ of spin $s$ is given by
\eqn\defcovspin{
\nabla_\mu \varphi_{(s)} = (\d_\mu - i s \omega_\mu)\varphi_{(s)}~.
}
In particular, on left- and right-moving spinors,
\eqn\covderspinors{
\nabla_\mu \psi_-= (\d_\mu - {i\over 2} \omega_\mu)\psi_-~,\qquad
\nabla_\mu \psi_+= (\d_\mu + {i\over 2} \omega_\mu)\psi_+~.
}
On one-forms, we have
\eqn\Doneform{\nabla_\mu X_1 = e^z_1\,  \nabla_\mu X_z~, \qquad \nabla_\mu X_{\b 1} = e^\bz_{\b 1}\,  \nabla_\mu X_\bz~,}
with $\nabla_\mu$ on the right-hand-side the Levi-Civita connection. (Note that $X_1$ and $X_{\b 1}$ in the frame basis are fields of spin $\pm 1$, respectively.)

Another important operator is the Lie derivative on a field of arbitrary spin along a (Killing) vector $K$:
\eqn\LieDeriv{
\CL_K \varphi_{(s)}= \left[K^\mu (\d_\mu-is\omega_\mu) + {is\over 2} \epsilon^{\mu\nu}\nabla_\mu K_\nu\right] \varphi_{(s)}~.
}
On can easily check that $\CL_K$ is metric-independent for any vector $K$  and that it reduces to the usual Lie derivative on forms for $s\in \Z$ whenever $K$ is Killing.

\appendix{B}{Some Useful Relations}
In the frame basis, the Killing spinor equation \KSEzeta\ reads
\eqn\KSEzetaii{
 D_1\zeta_- =0~, \qquad D_{\b 1}\zeta_- = \half\CH\zeta_+~, \qquad
 D_1  \zeta_+ = \half\t\CH \zeta_-~,  \qquad   D_{\b 1}\zeta_+ = 0~,
}
while \KSEzetat\ reads
\eqn\KSEzetatii{
 D_{ 1} \t\zeta_- =0~, \qquad D_{\b 1} \t\zeta_- = \half\t\CH\t\zeta_+~,
\qquad  D_{ 1} \t \zeta_+ = \half\CH\t\zeta_-~,  \qquad  D_{\b 1}\t\zeta_+ = 0~,
}
where the covariant derivative is defined as in \defDfullcov, with $r=\pm 1$ and $z=\t z=0$ for $\zeta, \t\zeta$, respectively. Let us define  the $R$-symmetry field strength
\eqn\defF{F_{1\b 1}\equiv  \d_1 A_{\b 1} - \d_{\b 1} A_1~.}
The following relations directly follow from \KSEzetaii:
\eqn\integraconds{\eqalign{
& 2 \zeta_- F_{1\b 1} + {i\over 4} \zeta_- (R- 2 \CH \t \CH) - i \zeta_+ \d_1 \CH =0\, , \cr
& 2 \zeta_+ F_{1\b 1} - {i\over 4} \zeta_+ (R- 2 \CH \t \CH) + i \zeta_- \d_{\b 1} \t \CH  =0 \, .
}}
Similarly, from \KSEzetatii\ we find
\eqn\integracondsii{\eqalign{
& -2 \t\zeta_- F_{1\b 1} + {i\over 4} \zeta_- (R- 2 \CH \t \CH) - i \t\zeta_+ \d_1 \t\CH =0\, , \cr
& -2 \t\zeta_+ F_{1\b 1} - {i\over 4} \zeta_+ (R- 2 \CH \t \CH) + i \t\zeta_- \d_{\b 1} \CH  =0 \, .
}}
Equations \integraconds, \integracondsii\ are very useful for the computations of section 6. These relations also imply that
\eqn\KonH{
K^\mu \d_\mu \CH=0~, \qquad K^\mu \d_\mu \t\CH=0~,
}
where $K^\mu$ is the Killing vector \defK.

\appendix{C}{Dimensional Reduction and Uplift to Three and Four dimensions}
Many of the two-dimensional supersymmetric backgrounds discussed in this paper can be obtained from  twisted dimensional reduction from three or four dimensions, and many of the two-dimensional supersymmetry transformations and Lagrangians of section 6 can be formally obtained by dimensional reduction of the three-dimensional results of \ClossetRU.  In this section, we briefly spell out this relation and discuss some interesting examples.  (We refer to the Appendix of \ClossetRU\ for a thorough discussion of the same dimensional reduction procedure from four to three dimensions.)

\subsec{Relation to Rigid Supersymmetry on Three-Manifolds}
Three-dimensional rigid supersymmetry for 3d $\CN=2$ theories with an $R$-symmetry was systematically studied in \ClossetRU. The three-dimensional Killing spinor equation reads \refs{\ClossetRU,\KlareGN}
\eqn\tdKSE{
(\nabla_M - i {\bf A}_M)\zeta   =  -\half {\bf H} \gamma_M \zeta - i {\bf V}_M\zeta -\half \epsilon_{MNR}{\bf V}^N \gamma^R\zeta~,
}
where $M, N, \cdots$ denotes three-dimensional coordinate indices and ${\bf H},  {\bf A}_M, {\bf V}_M$ denote the three-dimensional supergravity backgrounds fields.  Let us consider $\CM_3$  a fiber bundle over a two-manifold $\Sigma$, with metric
\eqn\tdmetric{
ds^2 = \eta^2 + g_{\mu\nu}^{\Sigma}(x)\, dx^\mu dx^\nu~, \qquad \eta = d\tau + c_\mu(x)\, dx^\mu~,
}
where $\mu, \nu$ run along $\Sigma$,  $g_{\mu\nu}^{\Sigma}$ is the two-dimensional metric, and $\tau$ is the coordinate along the fiber. It is convenient to choose a frame $E^{\bf A}$,
\eqn\frametd{
E^{\bf 1} = \eta~, \quad E^{\bf 2}= e^1~, \quad E^{\bf 3}= e^2~,
}
with $e^a$ ($a=1,2$) the two-dimensional frame, $g_{\mu\nu}^{\Sigma}=\delta_{ab}\,e^a_\mu e^b_\nu$. 
The three-dimensional gamma-matrices are similarly related to the two-dimensional ones by $\gamma^{\bf 1} = \gamma^3$,  $\gamma^{\bf 2} = \gamma^1$, $\gamma^{\bf 3} = \gamma^2$.
 Let us assume that we have a Killing spinor $\zeta$ which is $\tau$-independent in this frame.
One can check that the projection of the Killing spinor equation \tdKSE\  along $E^{\bf 1}=\eta$ is solved without imposing additional constraints on $\zeta$ if
\eqn\solveleg{
{\bf H} = {i\over 2} \epsilon^{\mu\nu}\d_\mu c_\nu~,\qquad  {\bf V}_{\bf 1}= {\bf A}_{\bf 1}~, \qquad {\bf V}_{a}=0~.
}
The remaining two legs of \tdKSE\ reproduce the two-dimensional Killing spinor equation \KSERi, 
\eqn\KSERii{
 (\nabla_\mu -i A_\mu) \zeta = -\half H \gamma_\mu \zeta +{i\over 2} G\gamma_\mu \gamma^3 \zeta~,
}
once we identify
\eqn\identtd{
H= 2 {\bf H}~, \qquad G= \eta^M {\bf A}_M = \eta^M {\bf V}_M~, \qquad A_\mu = {\bf A}_\mu -(\eta^M {\bf A}_M)  \eta_\mu~.
}
The same identifications also give a consistent reduction of the Killing spinor equation for the Killing spinor $\t\zeta$ of opposite $R$-charge. 

Let us denote by ${\bf C}_M$ the three-dimensional graviphoton, with ${\bf V}_M = -i \epsilon_{MNR}\d^N {\bf C}^R$. Due to \solveleg\ it can be taken along $\Sigma$ only. If we define $C_\mu^H = - c_\mu$ and $C_\mu^G= {\bf C}_\mu$, the two-dimensional graviphotons $C_\mu, \t C_\mu$ with field strengths \defHHtfromB\ are given by
\eqn\graviphotons{
C_\mu = C_\mu^{H}+ i C_\mu^{G}~, \qquad 
\t C_\mu = C_\mu^{H}- i C_\mu^{G}~.
}
 Finally, let us mention that momentum along $\tau$ gives rises to a real central charge $Z_{(i)}$, in addition to  the three-dimensional real central charge $Z_{(r)}$. One then defines the complex central charge
\eqn\defZ{
Z= Z_{(r)}+ i Z_{(i)}~, \qquad   \t Z= Z_{(r)}- i Z_{(i)}~.
}
The covariant derivative that appears on fields of definite charges $R, Z_{(r)}, Z_{(i)}$ after dimensional reduction is $D_\mu = \nabla_\mu - i r A_\mu - i z_{(r)} C_\mu^G + i z_{(i)} C_\mu^H$. 

By carefully performing this dimensional reduction, it is straightforward to derive the supersymmetry algebra \rigidsalg\ and the supersymmetry transformations for the general multiplet from the corresponding formulas in \ClossetRU.

\subsec{Uplift of Two-Dimensional Backgrounds to Higher Dimensions}
Conversely, it is easy to consider the higher-dimensional uplift of generic two-dimensional backgrounds.
Given such a background $(g^\Sigma_{\mu\nu}, A_\mu, C_\mu^H, C_\mu^G)$ with 
\eqn\defCHG{H= -i \epsilon^{\mu\nu}\d_\mu C_\nu^H~, \qquad
G = -i \epsilon^{\mu\nu}\d_\mu C_\nu^G~,}
we can directly write down a three-dimensional background
\eqn\tduplift{\eqalign{
& ds^2(\CM_3) = \eta^2 +  g_{\mu\nu}^{\Sigma}(x)\, dx^\mu dx^\nu~, \qquad \eta = d\tau - C_\mu^H(x)\, dx^\mu~,\cr
& {\bf H}^{(3d)}= \half H~, \qquad {\bf A}^{(3d)}= \eta G + A_\mu dx^\mu~, \qquad {\bf V}^{(3d)}= \eta G~,
}}
where $\CM_3$ is a circle bundle over $\Sigma$ with coordinates $(\tau, x^\mu)$. The background \tduplift\ preserves the same amount of supersymmetry as its two-dimensional reduction.%
\foot{Since \tduplift\ is a Seifert manifold it preserves at least two supercharges \ClossetRU.}
 Since the graviphoton $C_\mu^H$ appears as it does in \tduplift, a two-dimensional background must have a purely imaginary value of $H$ in order to admit an uplift to a three-dimensional background with real metric.

We can similarly uplift \tduplift\ to a four dimensional background \ClossetRU,
\eqn\fduplift{\eqalign{
& ds^2(\CM_4) = \eta^2+ \kappa^2 +  g_{\mu\nu}^{\Sigma}(x)\, dx^\mu dx^\nu~, \qquad \kappa = dy + C_\mu^G(x)\, dx^\mu~,\cr
& {\bf A}^{(4d)}= \half(\kappa H+ \eta G) + A_\mu dx^\mu~, \qquad\quad {\bf V}^{(4d)}=  \half(\kappa H+ \eta G)~,
}}
with $\eta$ defined in \tduplift.
 The resulting  four-manifold $\CM_4$ is (locally) a $T^2$ fibration over $\Sigma$ \DumitrescuHA, with  coordinates $\tau, y$ along the fiber.
This further uplift is allowed if and only if $G$ is also purely imaginary.

As a simple example, let us consider the maximally supersymmetric $S^2$ \spherefourQ, which reads (using coordinates $\theta, \varphi$)
\eqn\spherefourQii{
ds^2 = R_{S^2}^2 (d\theta^2 +\sin^2\theta d\varphi^2)~, \quad A_\mu =0~, \quad H = i{\lambdaa^2+\lambdaa^{-2}\over 2 R_{S^2}}~, \quad G= {\lambdaa^2-\lambdaa^{-2}\over 2 R_{S^2}}~,
}
with $\lambdaa\in \C$. To admit an uplift to three dimensions,  $H$ should be purely imaginary and therefore $\lambdaa$ should be either real or a pure phase. In the former case we can take $\lambdaa^2= \pm e^\alpha$, $\alpha \in \R$, and the three dimensional background \tduplift\ is
\eqn\tduexpli{\eqalign{
& ds^2 = {R_{S^3}^2\over 4} \left(\cosh^2\alpha (d\psi-\cos\theta d\varphi)^2 +  d\theta^2 + \sin^2\theta d\varphi^2 \right)~,\cr
& {\bf H}^{(3d)}= \pm i {\cosh\alpha\over R_{S^3}}~, \qquad {\bf A}^{(3d)}={\bf V}^{(3d)} = {\sinh{(2\alpha)}\over 2 R_{S^3}}   \, (d\psi -\cos\theta d\varphi)~,
}}
where we defined the coordinate $\psi =\pm {1\over  R_{S^2} \cosh \alpha}\tau$ and the radius $R_{S_3}= 2 R_{S^2}$.
This is the squashed three-sphere $S^3_b$ with  $SU(2)\times U(1)$ isometry  \ImamuraWG, with squashing parameter $b=\pm e^{\alpha}$.%
\foot{One can compare in particular to equation (5.11) of \ClossetRU.}
Note that in this case $G$ in \spherefourQii\ is real and therefore \tduexpli\ does not uplift further to four dimensions.
The second possibility is to take $\lambdaa = \pm e^{i \beta}$, $\beta \in [-{\pi\over 2},{\pi\over 2}]$. For $\beta\neq \pm {\pi\over 2}$,  the uplift of \spherefourQii\ is again to $S^3_b$ in three dimensions with squashing parameter $b=\pm e^{i\beta}$,
\eqn\tduexplii{\eqalign{
& ds^2 = {R_{S^3}^2\over 4} \left(\cos^2\beta (d\psi-\cos\theta d\varphi)^2 +  d\theta^2 + \sin^2\theta d\varphi^2 \right)~,\cr
& {\bf H}^{(3d)}= \pm i {\cos\beta\over R_{S^3}}~, \qquad {\bf A}^{(3d)}={\bf V}^{(3d)} = i{\sin{(2\beta)}\over 2 R_{S^3}}   \, (d\psi -\cos\theta d\varphi)~.
}}
In the limit $\beta= \pm {\pi\over 2}$, the background \spherefourQii\ uplifts instead to the $S^2 \times S^1$ background with maximal supersymmetry of \refs{\ImamuraUW,\ImamuraSU},
\eqn\tduexpliii{\eqalign{
& ds^2 =R_{S^1}^2 d u^2 +  R_{S^2}^2(d\theta^2 + \sin^2\theta d\varphi^2)~,\cr
&{\bf H}^{(3d)}= 0~, \qquad {\bf A}^{(3d)}={\bf V}^{(3d)} = \pm  i {R_{S^1}\over R_{S^2}} d u~.
}}
In the latter two cases, $G$ is purely imaginary and we can further uplift to four dimensions. In terms of the coordinates
\eqn\newcoord{
u={1\over R_{S^1}}(\cos\beta\, y+ \sin\beta\, \tau)~, \qquad
\psi=\pm{1\over R_{S^2}}(\sin\beta\, y- \cos\beta\, \tau)~, 
}
the four-dimensional uplift \fduplift\ of \spherefourQii\ with $\lambdaa^2 = \pm e^{i\beta}$ gives
\eqn\fdupliftii{\eqalign{
& ds^2=  R_{S^1}^2 du^2 + {R_{S^3}^2\over 4} \left( (d\psi-\cos\theta d\varphi)^2 +  d\theta^2 + \sin^2\theta d\varphi^2 \right)~,\cr
& {\bf A}^{(4d)}= {\bf V}^{(4d)}=  \pm {i\over R_{S^3}} du~,
}}
with $R_{S^3}= 2 R_{S^2}$, which is simply $S^3\times S^1$ with the round metric. Note that we obtain the same four-dimensional background for any value of $\lambdaa=\pm e^{i\beta}$, since $\beta$ is merely a rotation of the coordinates \newcoord. More generally, two-dimensional backgrounds related by an axial $R$-symmetry rotation uplift locally to the same four-dimensional geometry,  the axial $R$-symmetry being merely a frame rotation of the $T^2$ fiber. This is because the axial $R$-symmetry of 2d $\CN=(2,2)$ supersymmetry originates from the rotation symmetry of the $34$ plane when reducing 4d $\CN=1$ supersymmetry down to two dimensions.

Similar uplift formulas apply to the $U(1)$-isometric squashed $S^2$ background of section 4.3 and axial R-symmetry rotations thereof (with $\lambdaa$ a pure phase or real), leading generically to metrics with $U(1)^2$ isometry on the 3-sphere.

\appendix{D}{Embeddings into the General Multiplet}
In this Appendix, we spell  out the embedding of the supersymmetry multiplets discussed in sections 6.3 and 6.4 into the general multiplet $\CS$ with components \Scomponents.

\medskip

\item{1.)} {\it Chiral multiplet.} For a chiral multiplet $\Phi$ of charges $r, z, \t z$, we have the embedding
\eqn\embedChiral{\eqalign{
& C= \phi\, ,\quad  \chi_\pm = -\sqrt2 i \psi_\pm \, ,\quad \t\chi_\pm=0\, ,\quad M = -2 i F\, , \quad \t M =0\, , \cr
&  a_1= - i D_1\phi\, , \quad  a_{\b 1}= - i D_{\b 1}\phi \, , \quad \sigma= \zH \phi\, , \quad \t\sigma= \zHt\phi\, , \cr
& \lambda_\pm=0\, , \quad \t\lambda_\pm=0\, , \quad D= \left[{r\over 4} R - \half ( \t\CH z +\CH \t z)\right]\phi~.
}}

\item{2.)} {\it Antichiral multiplet.} For an antichiral multiplet $\t \Phi$ of charges $-r, -z, -\t z$, we have
\eqn\embedAntiChiral{\eqalign{
& C= \t\phi\, ,\quad  \chi_\pm = 0 \, ,\quad \t\chi_\pm=\sqrt2 i \t\psi_\pm\, ,\quad M = 0\, , \quad \t M = 2 i \t F\, , \cr
&  a_1= i D_1\t\phi\, , \quad  a_{\b 1}= i D_{\b 1}\t\phi \, , \quad \sigma= \zH \t\phi\, , \quad \t\sigma= \zHt\t\phi\, , \cr
& \lambda_\pm=0\, , \quad \t\lambda_\pm=0\, , \quad D= \left[{r\over 4} R - \half ( \t\CH z +\CH \t z)\right]\t\phi~.
}}

\item{3.)} {\it Twisted chiral multiplet.} The twisted chiral multiplet $\Omega$ has charges $r=z=\t z=0$ and embedding
\eqn\embedTwistedChiral{\eqalign{
& C= \omega\, ,\quad \chi_- = 0  \, , \quad  \chi_+ = -i\sqrt2  \t\eta_+  \, , \quad \t\chi_-=-i\sqrt2 \eta_-\, ,\quad \t\chi_+=0\, , \cr
& M = \t M= 0\, , \quad  a_1=  i D_1\omega\, , \quad  a_{\b 1}= - i D_{\b 1}\omega \, , \quad \sigma=0 \, , \quad \t\sigma= 2 i G\, , \cr
& \lambda_-=0\, , \quad \lambda_+= -2\sqrt2 D_{\b 1}\eta_-\, ,\quad  \t\lambda_-= 2\sqrt2 D_1 \t\eta_+\, , \quad \t\lambda_+=0~, \cr
& D= - 4 D_1 D_{\b 1} \omega\, .
}}

\item{4.)} {\it Twisted antichiral multiplet.} The twisted antichiral multiplet $\t\Omega$ has charges $r=z=\t z=0$ and embedding
\eqn\embedTwistedChiral{\eqalign{
& C= \t\omega\, ,\quad \chi_- = i\sqrt2 \t\eta_-  \, , \quad  \chi_+ =0  \, , \quad \t\chi_-=0\, ,\quad \t\chi_+= i\sqrt2  \eta_+\, , \cr
& M = \t M= 0\, , \quad  a_1=  -i D_1\t\omega\, , \quad  a_{\b 1}=  i D_{\b 1}\t\omega \, , \quad \sigma=-2 i \t G \, , \quad \t\sigma= 0\, , \cr
& \lambda_-=2\sqrt2 D_1 \eta_+\, , \quad \lambda_+=0 \, ,\quad  \t\lambda_-= 0\, , \quad \t\lambda_+= -2\sqrt2 D_{\b 1}\t \eta_-\, \cr
& D= - 4 D_1 D_{\b 1} \t\omega\, .
}}

\item{5.)} {\it Linear multiplet.} The linear multiplet $\CJ$ has charges $r=z=\t z=0$ and embedding
\eqn\embedLinear{\eqalign{
& C= J\, ,\quad  \chi_\pm = j_\pm \, ,\quad \t\chi_\pm=\t j_\pm\, ,\quad M =\t M =0\, , \cr
&  a_1= - j_1\, , \quad  a_{\b 1}= - j_{\b 1} \, , \quad \sigma= -K \, , \quad \t\sigma= -\t K\, , \cr
& \lambda_-= -2 i D_1 \t j_+ \, , \quad  \lambda_+= -2 i D_{\b 1} \t j_-\, , \quad 
\t \lambda_-= 2 i D_1  j_+ \, , \quad  \t\lambda_+= 2 i D_{\b 1}  j_-\, ,\cr
& D= - 4 D_1 D_{\b 1} J~,
}}
where $j_\mu$ is a conserved current, $\nabla_\mu j^\mu=0$.

\item{6.)} {\it Twisted linear multiplet.} The twisted linear multiplet $\h\CJ$ has charges $r=z=\t z=0$ and embedding
\eqn\embedLinear{\eqalign{
& C= \h J\, ,\quad  \chi_\pm = \h j_\pm \, ,\quad \t\chi_\pm=\t {\h j}_\pm\, ,\quad M =\h K\, , \quad \t M = \t{\h K}\, , \cr
&  a_1= \h j_1\, , \quad  a_{\b 1}= -\h j_{\b 1} \, , \quad \sigma= \t\sigma= 0\, , \quad \lambda_\pm = 0 \, , \quad \t\lambda_\pm = 0  \, , \quad D=0 \, ,        \cr
}}
where $\h j_\mu$ is a conserved current, $\nabla_\mu \h j^\mu=0$.

\appendix{E}{Semichirals Multiplets in Curved Space}
In flat space, a left semi-chiral multiplet ${\bf X}$ and a left semi-antichiral multiplet $\t {\bf X}$ are defined by the superspace constraints \BuscherUW
\eqn\defsemichiralX{
\t D_+ {\bf X} =0~, \qquad\qquad
D_+ \t {\bf X} =0~.
}
Similarly, a right semi-chiral multiplet ${\bf Y}$ and a right semi-antichiral multiplet $\t {\bf Y}$ satisfy
\eqn\defsemichiralX{
\t D_- {\bf Y} =0~, \qquad\qquad
D_- \t {\bf Y} =0~.
}
These multiplets have $4+4$ components.
In this Appendix, we briefly discuss  the semi-chiral multiplets in our formalism, which allows to discuss their coupling to curved space. We will not consider the gauging of these multiplets, which has  been investigated in flat space relatively recently \refs{\BuscherUW,\GatesVE, \LindstromHX, \CrichignoAA}.

\subsec{Semi-Chiral Multiplets Supersymmetry Transformations}
A left semi-chiral multiplet ${\bf X}$ of charges $r, z, \t z$ is a general multiplet with the single constraint $\t\chi_+=0$. It has components
\eqn\compoleftsemichiral{
{\bf X}= \big(X, \psi^L_-, \psi^L_+, \t\eta_- , F^L, v_{21}, \t\kappa^L, \t\rho_-\big)~,
}
which are embedded into the general multiplet \Scomponents\ as
\eqn\embedleftsemidChiral{\eqalign{
& C= X\, ,\quad  \chi_\pm = -\sqrt2 i \psi^L_\pm \, ,\quad \t\chi_- =- \sqrt2 i \t\eta_-\, , \quad \t\chi_+ =0\, ,
\cr &M = -2 i F^L\, , \quad \t M =0\, ,\quad  a_1= v_{21}\, , \quad  a_{\b 1}= - i D_{\b 1}X \, , \quad \sigma= z_\CH X\, , \quad \t\sigma= \t\kappa^L\, , \cr
& \lambda_-=\sqrt2  z_{\CH} \t\eta_-\, , \quad  \lambda_+ = -2\sqrt2 D_{\b 1} \t\eta_- \, , \quad \t\lambda_-=\t\rho_- \, , \quad \t\lambda_+=0\, , \cr 
&D= 2 i D_{\b 1} v_{21} - 2 D_1 D_{\b 1}X -\half z_{\CH} \t\kappa^L   \cr &\qquad
-\left(\t\CH  z_{\CH} - \half z_{\CH} \t z_{\CH} + 2 i r F_{1\b1} -{r\over 4}(R- 2 \CH\t\CH)\right)X~,
}}
where we used the notation \defzh.
Its susy variations are
\eqn\susyleftsemichir{\eqalign{
&\delta X = \sqrt2 (\zeta_+\psi^L_- - \zeta_-\psi^L_+)+\sqrt2 \t\zeta_+\t\eta_- \, ,\cr
&\delta \psi^L_- =\sqrt2 \zeta_- F^L -i \sqrt2 \t\zeta_- z_{\CH}  X  +i \sqrt2\t\zeta_+ \left(D_1 X+iv_{21}\right)~,\cr
&\delta \psi^L_+ = \sqrt2 \zeta_+ F^L -{i\over\sqrt2}\t\zeta_+ \left(\t \kappa^L +\t z_\CH X \right)  + 2i\sqrt2\t\zeta_- D_{\b 1} X~,\cr
&\delta \t\eta_- = - {i\over \sqrt2}\zeta_- \left(\t\kappa^L  - \t z_\CH X \right)  + i\sqrt2\zeta_+ \left(D_1 X- i v_{21}\right)~, \cr
&\delta F^L = -i \t\zeta_+\t\rho_- +i\sqrt2  \t z_\CH \t\zeta_+ \psi^L_- - i \sqrt2 z_\CH\t\zeta_- \psi^L_+  
+ 2 i\sqrt2 \t\zeta_+ D_1 \psi^L_+ - 2 i\sqrt2 \t\zeta_- D_{\b 1} \psi^L_-~,\cr
&\delta v_{21} = - i  \zeta_- \t\rho_- -i\sqrt2 \t\zeta_- z_\CH \eta_- - i\sqrt2 D_1  \left( \zeta_+\psi^L_- - \zeta_-\psi^L_+ - \t\zeta_+\t\eta_- \right) ~,   \cr
&\delta \t\kappa^L =  4\sqrt2 \t\zeta_- D_{\b1} \t\eta_- - 2\zeta_+ \t\rho_- + \sqrt2\t z_\CH \left( \zeta_+\psi^L_- - \zeta_-\psi^L_+ - \t\zeta_+\t\eta_- \right)~, \cr 
&\delta \t\rho_- =
 -i \t\zeta_- \left(  4 i D_{\b1}v_{21}- 4 D_1D_{\b 1}X + \CH \t\kappa^L - z_\CH \t\kappa^L   \right)\cr
&  \quad  +i \t\zeta_- \left(\t\CH z_\CH - z_\CH\t z_\CH + 2 i r F_{1\b1} -{r\over4}(R- 2\CH\t\CH)\right)X  - 2 i \t\zeta_+ \left(D_1 \t\kappa^L - i \t z_\CH v_{21}  \right)~.
}}
Note that, in the present formalism, these multiplet would be best called semi-ugly multiplets.
Similarly, a left semi-antichiral multiplet ${\bf X}$ of charges $-r, - z, -\t z$ corresponds to $\chi_+=0$ in the general multiplet. It has components
\eqn\compoleftsemiantchiral{
\t{\bf X}= (\t X, \eta_- , \t\psi^L_-, \t\psi^L_+, \t F^L, v_{11}, \kappa^L, \rho_-)~,
}
with embedding into the general multiplet
\eqn\embedleftsemidAntiChiral{\eqalign{
& C= \t X\, ,\quad \chi_- = \sqrt2 i \eta_-\, , \quad \chi_+ =0\, , \quad \t\chi_\pm = \sqrt2 i \t\psi^L_\pm 
\cr &M =0\, , \quad \t M =2 i \t F^L\, ,\quad  a_1= v_{11}\, , \quad  a_{\b 1}= i D_{\b 1}\t X \, , \quad \sigma= \kappa^L\, , \quad \t\sigma= \t z_\CH \t X\, , \cr
& \lambda_-=\rho_-\, , \quad  \lambda_+ = 0 \, , \quad \t\lambda_-=\sqrt2  \t z_{\CH} \eta_- \, , \quad \t\lambda_+=- 2\sqrt2 D_{\b1} \eta_-\, , \cr 
&D= -2 i D_{\b 1} v_{11} - 2 D_1 D_{\b 1}\t X -\half \t z_{\CH} \kappa^L  \cr &\qquad -\left(\CH \t z_{\CH} - \half z_{\CH} \t z_{\CH} - 2 i r F_{1\b1} -{r\over 4}(R- 2 \CH\t\CH)\right)\t X~.
}}
Its supersymmetry transformations are
\eqn\susyleftsemiantichir{\eqalign{
&\delta \t X = -\sqrt2 \zeta_+\eta_-  -\sqrt2 (\t\zeta_+\t\psi^L_- - \t\zeta_-\t\psi^L_+)\, ,\cr
&\delta \eta_- =  {i\over \sqrt2}\t\zeta_- \left(\kappa^L  -  z_\CH \t X \right)  - i\sqrt2\t\zeta_+ \left(D_1 \t X+ i v_{11}\right)~, \cr
&\delta \t\psi^L_- =\sqrt2\t\zeta_- \t F^L +i \sqrt2 \zeta_- \t z_{\CH}\t  X  -i \sqrt2\zeta_+ \left(D_1 \t X-i v_{11}\right)~,\cr
&\delta \t\psi^L_+ = \sqrt2 \t\zeta_+\t F^L +{i\over\sqrt2}\zeta_+ \left( \kappa^L + z_\CH \t X \right)  +-2i\sqrt2\zeta_- D_{\b 1} \t X~,\cr
&\delta \t F^L = -i \zeta_+\rho_- +i\sqrt2   z_\CH \zeta_+ \t\psi^L_- - i \sqrt2 \t z_\CH\zeta_- \t\psi^L_+  
+ 2 i \sqrt2\zeta_+ D_1 \t\psi^L_+ - 2 i \sqrt2\zeta_- D_{\b 1} \t\psi^L_-~,\cr
&\delta v_{11} = -i  \t\zeta_- \rho_- -i\sqrt2 \zeta_-\t z_\CH \t\eta_- - i\sqrt2 D_1  \left( \t\zeta_+\t\psi^L_- - \t\zeta_-\t\psi^L_+ - \zeta_+\eta_- \right) ~,   \cr
&\delta \kappa^L =  -4\sqrt2 \zeta_- D_{\b1} \eta_- + 2\t\zeta_+ \rho_- - \sqrt2 z_\CH \left( \t\zeta_+\t\psi^L_- - \t\zeta_-\t\psi^L_+ - \zeta_+\eta_- \right)~, \cr 
&\delta \rho_- =
 i \zeta_- \left(  -4 i D_{\b1}v_{11}- 4 D_1D_{\b 1}\t X + \t\CH \kappa^L - \t z_\CH \kappa^L   \right)\cr
&  \quad  -i \zeta_- \left(\CH \t z_\CH - z_\CH\t z_\CH - 2 i r F_{1\b1} -{r\over4}(R- 2\CH\t\CH)\right)\t X  + 2 i \zeta_+ \left(D_1 \kappa^L + i  z_\CH v_{11}  \right)~.
}}

A right semi-chiral multiplet  ${\bf Y}$ of charges $r, z, \t z$ is a general multiplet with the constraint $\t\chi_-=0$. It has components
\eqn\comporightsemichiral{
{\bf Y}= (Y, \psi^R_-, \psi^R_+, \t\eta_+ , F^R, v_{2\b 1}, \kappa^R, \t\rho_+)~,
}
and embedding
\eqn\embedrightsemidChiral{\eqalign{
& C= Y\, ,\quad  \chi_\pm = -\sqrt2 i \psi^R_\pm \, ,\quad \t\chi_- =0\, , \quad \t\chi_+ =- \sqrt2 i \t\eta_+\, ,
\cr &M = -2 i F^R\, , \quad \t M =0\, ,\quad  a_1= -i D_1 Y\, , \quad  a_{\b 1}= v_{2\b1} \, , \quad \sigma= \kappa^R\, , \quad \t\sigma= \t z_\CH Y\, , \cr
& \lambda_-= -2\sqrt2 D_{ 1} \t\eta_+ \, , \quad  \lambda_+ =\sqrt2\t z_{\CH} \t\eta_+\, , \quad \t\lambda_-=0 \, , \quad \t\lambda_+=\t\rho_+\, , \cr 
&D= 2 i D_{ 1} v_{2\b 1} - 2 D_{\b 1} D_{1}Y -\half \t z_{\CH} \kappa^R \cr &\qquad  -\left(\CH \t z_{\CH} - \half z_{\CH} \t z_{\CH} - 2 i r F_{1\b1} -{r\over 4}(R- 2 \CH\t\CH)\right)Y~.
}}
Its supersymmetry transformations are
\eqn\susyrightsemichir{\eqalign{
&\delta Y = \sqrt2 (\zeta_+\psi^R_- - \zeta_-\psi^R_+)-\sqrt2 \t\zeta_-\t\eta_+ \, ,\cr
&\delta \psi^R_- =\sqrt2 \zeta_- F^R -{i\over \sqrt2} \t\zeta_- (\kappa^R+z_{\CH}  Y)  +2i \sqrt2\t\zeta_+ D_1 Y~,\cr
&\delta \psi^R_+ = \sqrt2 \zeta_+ F^R -i\sqrt2\t\zeta_+\t z_\CH Y  + i\sqrt2\t\zeta_- (D_{\b 1} Y+ i v_{2\b1})~,\cr
&\delta \t\eta_+ = - {i\over \sqrt2}\zeta_+ \left(\t\kappa^R  -  z_\CH Y \right)  + i\sqrt2\zeta_- \left(D_{\b 1} Y- i v_{2\b1}\right)~, \cr
&\delta F^R = i \t\zeta_-\t\rho_+ +i\sqrt2  \t z_\CH \t\zeta_+ \psi^R_- - i \sqrt2 z_\CH\t\zeta_- \psi^R_+  
+ 2 i\sqrt2 \t\zeta_+ D_1 \psi^R_+ - 2 i\sqrt2 \t\zeta_- D_{\b 1} \psi^R_-~,\cr
&\delta v_{21} = i  \zeta_+ \t\rho_+ +i\sqrt2 \t\zeta_+ \t z_\CH \t\eta_+ - i\sqrt2 D_{\b 1}  \left( \zeta_+\psi^R_- - \zeta_-\psi^R_+ + \t\zeta_-\t\eta_+ \right) ~,   \cr
&\delta \t\kappa^R =  -4\sqrt2 \t\zeta_+ D_{1} \t\eta_+ + 2\zeta_- \t\rho_+ + \sqrt2 z_\CH \left( \zeta_+\psi^R_- - \zeta_-\psi^R_+ + \t\zeta_-\t\eta_+ \right)~, \cr 
&\delta \t\rho_+ =
 -i \t\zeta_+ \left(  4 i D_{1}v_{2\b 1}- 4 D_{\b1}D_{1}Y + \t\CH \kappa^R - \t z_\CH \kappa^R   \right)\cr
&  \quad  +i \t\zeta_+ \left(\CH \t z_\CH - z_\CH\t z_\CH - 2 i r F_{1\b1} -{r\over4}(R- 2\CH\t\CH)\right)Y  - 2 i \t\zeta_- \left(D_{\b 1} \kappa^R - i  z_\CH v_{2\b 1}  \right)~.
}}
Similarly, the right semi-antichiral of charge $-r, -z, -\t z$ corresponds to the constraint $\chi_-=0$. It has components
\eqn\comporightsemichiral{
\t{\bf Y}= (\t Y, \eta_+, \t\psi^R_-, \t\psi^R_+ , \t F^R, v_{1\b 1}, \t\kappa^R, \rho_+)~,
}
with embedding
\eqn\embedrightsemidChiral{\eqalign{
& C= \t Y\, ,\quad \chi_- =0\, , \quad \t\chi_+ = \sqrt2 i \eta_+\, ,\quad  \t\chi_\pm = \sqrt2 i \t\psi^R_\pm
\cr &M = 0\, , \quad \t M =2 i\t F^R\, ,\quad  a_1= i D_1 \t Y\, , \quad  a_{\b 1}= v_{1\b1} \, , \quad \sigma= z_\CH \t Y\, , \quad \t\sigma= \t \kappa^R\, , \cr
& \lambda_-=0 \, , \quad \lambda_+=\rho_+\, ,\quad 
 \t\lambda_-= -2\sqrt2 D_{ 1} \eta_+ \, , \quad  \t\lambda_+ =\sqrt2 z_{\CH} \eta_+\, ,  \cr 
&D=- 2 i D_{ 1} v_{1\b 1} - 2 D_{\b 1} D_{1}\t Y -\half  z_{\CH} \t\kappa^R   \cr &\qquad
-\left(\t\CH  z_{\CH} - \half z_{\CH} \t z_{\CH} + 2 i r F_{1\b1} -{r\over 4}(R- 2 \CH\t\CH)\right)\t Y~.
}}
Its supersymmetry transformations are left as an exercise.

\subsec{Supersymmetric Lagrangians}
Kinetic Lagrangians for the semi-chiral multiplets are easily obtained by extracting the $D$-term of
\eqn\Lagsemichiral{
\t{\bf X} {\bf X}+ \t{\bf Y}{\bf Y} + {1\over \alpha} \big( \t{\bf X} {\bf Y}+ \t{\bf Y}{\bf X}  \big)~,
}
with $\alpha \geq 1$ a free parameter. The resulting expressions are a bit long and we leave them as an exercise for the interested reader. (One just needs the product formula \productrulestwodim.) Note that we need to consider both left and right semi-chiral multiplets simultaneously. The theory of a single left semi-chiral multiplet with Lagrangian $\t{\bf X} {\bf X}\big|_D$ has no interesting dynamics.

\listrefs

\end